\documentclass[aps,prl,preprint,groupedaddress]{revtex4-1}
\usepackage{graphics}
\usepackage{graphicx}
\usepackage{amssymb, amsmath}
\usepackage{amsthm}
\usepackage{array}
\usepackage[T1]{fontenc}

\newtheorem{mydef}{Definition}

\begin{document}

% Use the \preprint command to place your local institutional report
% number in the upper righthand corner of the title page in preprint mode.
% Multiple \preprint commands are allowed.
% Use the 'preprintnumbers' class option to override journal defaults
% to display numbers if necessary
%\preprint{}

%Title of paper
\title{Learning from Complex Systems: On the Roles of Entropy and Fisher Information in Pairwise Isotropic Gaussian Markov Random Fields}

% repeat the \author .. \affiliation  etc. as needed
% \email, \thanks, \homepage, \altaffiliation all apply to the current
% author. Explanatory text should go in the []'s, actual e-mail
% address or url should go in the {}'s for \email and \homepage.
% Please use the appropriate macro foreach each type of information

% \affiliation command applies to all authors since the last
% \affiliation command. The \affiliation command should follow the
% other information
% \affiliation can be followed by \email, \homepage, \thanks as well.
\author{Alexandre L. M. Levada}
\email[]{alexandre@dc.ufscar.br}
\homepage[]{http://www.dc.ufscar.br/~alexandre}
%\thanks{}
%\altaffiliation{}
\affiliation{Universidade Federal de S\~ao Carlos, SP, Brazil}

%Collaboration name if desired (requires use of superscriptaddress
%option in \documentclass). \noaffiliation is required (may also be
%used with the \author command).
%\collaboration can be followed by \email, \homepage, \thanks as well.
%\collaboration{}
%\noaffiliation

\date{\today}

\begin{abstract}
%% Text of abstract
Markov Random Field models are powerful tools for the study of complex systems. However, little is known about how the interactions between the elements of such systems are encoded, especially from an information-theoretic perspective. In this paper, our goal is to enlight the connection between Fisher information, Shannon entropy, information geometry and the behavior of complex systems modeled by isotropic pairwise Gaussian Markov random fields. We propose analytical expressions to compute local and global versions of these measures using Besag's pseudo-likelihood function, characterizing the system's behavior through its \emph{Fisher curve}, a parametric trajectory accross the information space that provides a geometric representation for the study of complex systems. Computational experiments show how the proposed tools can be useful in extrating relevant information from complex patterns. The obtained results quantify and support our main conclusion, which is: in terms of information, moving towards higher entropy states (A --> B) is different from moving towards lower entropy states (B --> A), since the \emph{Fisher curves} are not the same given a natural orientation (the direction of time).
\end{abstract}

% insert suggested PACS numbers in braces on next line
% insert suggested keywords - APS authors don't need to do this
%\keywords{}

%\maketitle must follow title, authors, abstract, \pacs, and \keywords
\maketitle

% body of paper here - Use proper section commands
% References should be done using the \cite, \ref, and \label commands
\section{Introduction}

With the increasing value of information in modern society and the massive volume of digital data that is available, there is an urgent need of developing novel methodologies for data filtering and analysis in complex systems. In this scenario, the notion of what is informative or not is a top priority. Sometimes, patterns that at first may appear to be locally irrelevant may turn out to be extremely informative in a more global perspective. In complex systems, this is a direct consequence of the intricate non-linear relationship between the pieces of data along different locations and scales.

Within this context, information theoretic measures play a fundamental role in a huge variety of applications once they represent statistical knowledge in a sistematic, elegant and formal framework. Since the first works of Shannon \cite{shannon1949}, and later with many other generalizations \cite{renyi1961, tsallis1988, bashkirov2006}, the concept of entropy has been adapted and successfully applied to almost every field of science, among which we can cite physics \cite{jaynes1957}, mathematics \cite{grad1961, adler1965, goodwyn1972}, economics \cite{samuelson1972} and fundamentally, information theory \cite{costa1983, cover1991, cover1994}. Similarly, the concept of Fisher information \cite{Frieden2004, Frieden2006} has been shown to reveal important properties of statistical procedures, from lower bounds on estimation methods \cite{Lehmann1983, Bickel1991, Casella2002} to information geometry \cite{Amari, Kass1989}. Roughly speaking, Fisher information can be thought as the likelihood analog of entropy, which is a probability-based measure of uncertainty.

In general, classical statistical inference is focused on capturing information about location and dispersion of unknown parameters of a given family of distribution and studying how this information is related to uncertainty in estimation procedures. In typical situations, exponential family of distributions and independence hypothesis (independent random variables) are often assumed, giving the likelihood function a series of desirable mathematical properties \cite{Lehmann1983, Bickel1991, Casella2002}.

Although mathematically convenient for many problems, in complex systems modeling, independence assumption is not reasonable because much of the information is somehow encoded in the relations between the random variables \cite{Anandkumar2009, Villegas}. In order to overcome this limitation, Markov Random Field (MRF) models appear as a natural generalization of the classical approach by the replacement of the independence assumption by a more realistic conditional independence assumption. Basically, in every MRF, knowledge of a finite-support neighborhood aroung a given variable isolates it from all the remaining variables. A further simplification consists in considering a pairwise interaction model, constraining the size of the maximum clique to be two (in other words, the model captures only binary relationships). Moreover, if the MRF model is isotropic, which means that the parameter controlling the interactions between neighboring variables is invariant to change in the directions, all the information regarding the spatial dependence structure of the system is conveyed by a single parameter, from now on denoted by $\beta$ (or simply, the inverse temperature).

In this paper, we assume an isotropic pairwise Gaussian Markov Random Field (GMRF) model \cite{Moura1992, Moura1997}, also known as auto-normal model or conditional auto-regressive model \cite{Besag1974,Besag1975}. Basically, the question that motivated this work and we are trying to elucidate here is: \emph{What kind of information is encoded by the $\beta$ parameter in such a model?} We want to know how this parameter, and as a consequence, the whole spatial dependence structure of a complex system modelled by a Gaussian Markov random field, is related to both local and global information theoretic measures, more precisely the observed and expected Fisher information as well as self-information and Shannon entropy.

In searching for answers for our fundamental question, investigations led us to an exact expression for the asymptotic variance of the maximum pseudo-likelihood (MPL) estimator of $\beta$ in an isotropic pairwise GMRF model, suggesting that asymptotic efficiency is not granted. In the context of statistical data analysis, Fisher information plays a central role in providing tools and insights for modeling the interactions between complex systems and their components. The advantage of MRF models over the traditional statistical ones is that MRF's take into account the dependence between pieces of information as a function of the system's temperature, which may even be variable along the time. Briefly speaking, this investigation aims to explore ways to measure and quantify distances between complex systems operating in different thermodynamical conditions. By analyzing and comparing the behavior of local patterns observed throughout the system (defined over a regular 2D lattice), it is possible to measure how informative are those patterns for a given inverse temperature, or simply $\beta$ (which encodes the expected global behavior). 

The remaining of the paper is organized as follows: Section 2 discusses a technique for $\beta$ estimation called maximum pseudo-likelihood (MPL) and provides derivations for the observed Fisher information in an isotropic pairwise GMRF model. Intuitive interpretations for the two versions of this measure are discussed. In Section 3 we derive analytical expressions for the computation of the expected Fisher information. In Section 4 an expression for the global entropy in a GMRF model is shown. The results suggest a connection between maximum pseudo-likelihood and minimum entropy criteria in GMRF's. Section 5 discusses the asymptotic variance of $\beta$'s maximum pseudo-likelihood estimator. In Section 6 the definition of \emph{Fisher curve} of a system as a parametric trajectory in the information space is proposed. Section 7 shows the experimental setup. Computational simulations with both Markov Chain Monte Carlo algorithms and real data were conducted, showing the effectiveness of the proposed tools in extracting relevant information from complex systems. Finally, Section 8 presents our conclusions, final remarks and possibilities for future works.

\section{Fisher Information in Isotropic Pairwise GMRF's}

The remarkable Hammersley-Clifford theorem \cite{hammersley1971} states the equivalence between Gibbs Random Fields (GRF) and Markov Random Fields (MRF), which implies that any MRF can be defined either in terms of a global (joint Gibbs distribution) or a local (set of local conditional density functions) model. For our purposes, we will choose the later representation.

\begin{mydef}
An isotropic pairwise Gaussian Markov random field regarding a local neighborhood system $\eta_{i}$ defined on a lattice $S=\left\{ s_{1}, s_{2}, \ldots , s_{n} \right\}$ is completely characterized by a set of $n$ local conditional density functions $p( x_{i} | \eta_{i}, \vec{\theta} )$, given by:
\end{mydef}

\begin{equation}
	p\left( x_{i} | \eta_{i}, \vec{\theta} \right) = \frac{1}{\sqrt{2\pi}\sigma}exp\left\{-\frac{1}{2\sigma^{2}} \left[ x_{i} - \mu - \beta \sum_{j \in \eta_{i}} \left( x_{j} - \mu \right) \right]^{2} \right\}
	\label{eq:GMRF}
\end{equation} with $\vec{\theta} = (\mu, \sigma^{2}, \beta)$, where $\mu$ and $\sigma^{2}$ are the expected value and the variance of the random variables, and $\beta = 1/T$ is the parameter that controls the interaction between the variables (inverse temperature). Note that, for $\beta = 0$, the model degenerates to the usual Gaussian distribution. From an information geometry perspective \cite{Amari, Kass1989}, it means that we are constrained to a sub-manifold within the Riemmanian manifold of probability distributions, where the natural Riemmanian metric (tensor) is given by the Fisher information. It has been shown that the geometric structure of exponential family distributions exhibit constant curvature. However, little is known about information geometry on more general statistical models, such as GMRF's. For $\beta > 0$, some degree of correlation between the observations is expected, making the interactions grow stronger. Typical choices for $\eta_{i}$ are the first and second order non-causal neighborhood systems, defined by the sets of 4 and 8 nearest neighbors, respectively.

%%%%%%%%%%%%%%%%%%%%%%%%%%%%%%%%%%%%%%%%%%

\subsection{Maximum Pseudo-Likelihood Estimation}

Maximum likelihood estimation is intractable in MRF parameter estimation due to the existence of the partition function in the joint Gibbs distribution. An alternative, proposed by Besag \cite{Besag1974}, is maximum pseudo-likelihood estimation, which is based on the conditional independence principle. The pseudo-likelihood function is defined as the product of the LCDF's for all the $n$ variables of the system, modeled as a random field.

\begin{mydef}
Let an isotropic pairwise GMRF be defined on a lattice $S=\left\{ s_{1}, s_{2}, \ldots , s_{n} \right\}$ with a neighborhood system $\eta_{i}$. Assuming that $\mathbf{X^{(t)}}=\{x_{1}^{(t)}, x_{2}^{(t)}, \ldots, x_{n}^{(t)} \}$ denotes the set corresponding to the observations at time $t$, the pseudo-likelihood function of the model is defined by:
\end{mydef}

\begin{equation}
	L\left(\vec{\theta}; \mathbf{X}^{(t)}\right) = \prod_{i=1}^{n}p( x_{i} | \eta_{i}, \vec{\theta} )
	\label{eq:PL}
\end{equation}

Note that the pseudo-likelihood function is a function of the parameters. For better mathematical tractability, it is usual to take the logarithm of $L(\vec{\theta}; \mathbf{X}^{(t)})$. Plugging equation \eqref{eq:GMRF} into equation \eqref{eq:PL} and taking the logarithm, leads to:

\begin{equation}
	log~L\left(\vec{\theta}; \mathbf{X}^{(t)} \right) = -\frac{n}{2}log\left( 2\pi\sigma^{2} \right) -\frac{1}{2\sigma^{2}}\sum_{i=1}^{n}\left[ x_{i} - \mu - \beta\sum_{j \in \eta_i}\left( x_{j} - \mu \right) \right]^{2}
	\label{eq:GMRF_PL} 
\end{equation}

By differentiating equation \eqref{eq:GMRF_PL} with respect to each parameter and properly solving the pseudo-likelihood equations we obtain the following maximum pseudo-likelihood estimators for the parameters $\mu$, $\sigma^{2}$ and $\beta$:

\begin{equation}
	\hat{\beta}_{MPL} = \frac{\displaystyle\sum_{i=1}^{n}\left[\left( x_{i} - \mu \right)\displaystyle\sum_{j \in \eta_i}\left(x_{j} - \mu \right)\right]}{\displaystyle\sum_{i=1}^{n}\left[ \displaystyle\sum_{j \in \eta_i}\left( x_{j} - \mu  \right) \right]^{2}}
	\label{eq:BetaMPL}
\end{equation}
\vspace{0.2cm}

\begin{equation}
	\hat{\mu}_{MPL} = \frac{1}{n \left( 1 - k\beta \right)}\displaystyle\sum_{i=1}^{n}\left( x_{i} - \beta\displaystyle\sum_{j \in \eta_i}x_{j} \right)
	\label{eq:MuMPL}
\end{equation}	
\vspace{0.2cm}

\begin{equation}
	\hat{\sigma}_{MPL}^{2} = \frac{1}{n}\displaystyle\sum_{i=1}^{n}\left[ x_{i} - \mu - \beta\sum_{j \in \eta_i}\left( x_{j} - \mu \right) \right]^{2}
	\label{eq:SigmaMPL} 
\end{equation} 
\vspace{0.2cm}

\noindent where $k$ denotes the cardinality of the non-causal neighborhood set $\eta_{i}$. Note that if $\beta = 0$, the MPL estimators of both $\mu$ and $\sigma^{2}$ become the widely known sample mean and sample variance.

Since the cardinality of the neighborhood system, $k = |\eta_{i}|$, is spatially invariant (we are assuming a regular neighborhood system) and each variable is dependent on a fixed number of neighbors on a lattice, $\hat{\beta}_{MPL}$ can be rewritten in terms of cross covariances:
%(note that despite a constant factor the numerator and denominator of $\hat{\beta}$ are a sum of cross-covariances)

\begin{equation}
	\hat{\beta}_{MPL} = \frac{\displaystyle\sum_{j \in \eta_i}\hat{\sigma}_{ij}}{\displaystyle\sum_{j \in \eta_i}\displaystyle\sum_{k \in \eta_i}\hat{\sigma}_{jk}} 
	\label{eq:BetaMPL2}
\end{equation} where $\sigma_{ij}$ denotes the sample covariance between the central variable $x_{i}$ and $x_{j} \in \eta_{i}$. Similarly, $\sigma_{jk}$ denotes the sample covariance between two variables belonging to the neighbohood system $\eta_{i}$ (the definition of the neighborhood system $\eta_{i}$ does not include the the location $s_{i}$).

\subsection{Fisher information of spatial dependence parameters}

Basically, Fisher information measures the amount of information a sample conveys about an unknown parameter. It can be thought as the likelihood analog of entropy, which is a probability-based measure of uncertainty. Often, when we are dealing with independent and identically distributed (i.i.d) random variables, the computation of the global Fisher Infomation presented in a random sample $\mathbf{X}^{(t)} = \{ x_{1}^{(t)}, x_{2}^{(t)}, \ldots, x_{n}^{(t)} \}$ is quite straighforward, since each observation $x_{i}$, $i = 1,2,\ldots,n$, brings exactly the same amout of information (when we are dealing with independent samples, the superscript $t$ is usually supressed since the underlying dependence struture does not change through time). However, this is not true for spatial dependence parameters in MRF's, since different configuration patterns ($x_{i} \cup \eta_{i}$) provide distinct contributions to the local observed Fisher information, which can be used to derive a reasonable approximation to the global Fisher information \cite{Efron1978}. 

\subsection{The Information Equality} 

It is widely known from statistical inference theory that information equality holds in case of independent observations in the exponential family \cite{Lehmann1983, Bickel1991, Casella2002}. In other words, we can compute the Fisher information of a random sample regarding a parameter of interest $\theta$ by:

\begin{equation}
	I\left(\theta; \mathbf{X}^{(t)} \right) = E\left[ \left( \frac{\partial}{\partial \theta} log L\left(\theta; \mathbf{X}^{(t)} \right) \right)^{2} \right] = - E\left[ \frac{\partial^{2}}{\partial \theta^{2}} log L\left(\theta; \mathbf{X}^{(t)} \right) \right]
\end{equation} where $L\left(\theta; \mathbf{X}^{(t)} \right)$ denotes the likelihood function at a time instant $t$. In our investigations, to avoid the joint Gibbs distribution, often intractable due to the presence of the partition function (global Gibbs field), we replace the usual likelihood function by Besag's pseudo-likelihood function and then we work with the local model instead (local Markov field).

However, given the intrinsic spatial dependence struture of Gaussian Markov random field models, information equilibrium is not a natural condition. As we will discuss later, in general, information equality fails. Thus, in a GMRF model we have to consider two kinds of Fisher information, from now on denoted by type-I (due to the first derivative of the pseudo-likelihood function) and type-II (due to the second derivative of the pseudo-likelihood function). Eventually, when certain conditions are satisfied, these two values of information will converge to a unique bound. Essentially, $\beta$ is the parameter responsible to control whether both forms of information converge or diverge. Knowing the role of $\beta$ (inverse temperature) in a GMRF model, it is expected that for $\beta = 0$ (or $T \rightarrow \infty$) information equilibrium prevails. In fact, we will see in the following sections that as $\beta$ deviates from zero (and long-term correlations start to emerge), the divergence between the two kinds of information increases.

\subsection{Observed Fisher information} 

In order to quantify the amount of information conveyed by a local configuration pattern in a complex system, the concept of observed Fisher information must be defined.  

\begin{mydef}
Consider a MRF defined on a lattice $S=\left\{ s_{1}, s_{2}, \ldots , s_{n} \right\}$ with a neighborhood system $\eta_{i}$. The type-I local observed Fisher information for the observation $x_{i}$ regarding the spatial dependence parameter $\beta$ is defined in terms of its local conditional density function as:
\end{mydef}

\begin{equation}
	\phi_{\beta}(x_{i}) = \left[ \frac{\partial}{\partial\beta} log~p\left( x_{i} | \eta_{i}, \vec{\theta} \right) \right]^{2}	
\label{eq:phi_local} 
\end{equation}
\vspace{0.2cm}

Hence, for an isotropic pairwise GMRF model, the type-I local observed Fisher information regarding $\beta$ for the observation $x_{i}$ is given by:

\begin{align}
	\phi_{\beta}(x_{i}) & = \frac{1}{\sigma^{4}}\left\{ \left[ x_{i} - \mu - \beta \sum_{j \in \eta_{i}}\left( x_{j} - \mu \right) \right]\left[ \sum_{j \in \eta_{i}}\left( x_{j} - \mu \right) \right]\right\}^{2} \nonumber \\
	& = \frac{1}{\sigma^{4}}\left[ \sum_{j \in \eta_{i}}\left(x_{i} - \mu \right) \left(x_{j} - \mu \right) - \beta \sum_{j \in \eta_{i}}\sum_{k \in \eta_{i}}\left(x_{j} - \mu \right) \left(x_{k} - \mu \right)  \right]^{2}
\label{eq:phi_local_GMRF} 
\end{align}
\vspace{0.2cm}

\begin{mydef}
Consider a MRF defined on a lattice $S=\left\{ s_{1}, s_{2}, \ldots , s_{n} \right\}$ with a neighborhood system $\eta_{i}$. The type-II local observed Fisher information for the observation $x_{i}$ regarding the spatial dependence parameter $\beta$ is defined in terms of its local conditional density function as:
\end{mydef}

\begin{equation}
	\psi_{\beta}(x_{i}) = -\frac{\partial^{2}}{\partial\beta^{2}} log~p\left( x_{i} | \eta_{i}, \vec{\theta} \right)	
\label{eq:psi_local} 
\end{equation}
\vspace{0.2cm}

In case of an isotropic pairwise GMRF model, the type-II local observed Fisher information regarding $\beta$ for the observation $x_{i}$ is given by:

\begin{align}
	\phi_{\beta}(x_{i}) & = \frac{1}{\sigma^{2}}\left[ \sum_{j \in \eta_{i}}\sum_{k \in \eta_{i}}\left(x_{j} - \mu \right) \left(x_{k} - \mu \right) \right] 
\label{eq:psi_local_GMRF} 
\end{align}

Note that $\phi_{\beta}(x_{i})$ does not depend on $x_{i}$, only on the neighborhood system $\eta_{i}$.

\begin{mydef}
Consider a MRF defined on a lattice $S=\left\{ s_{1}, s_{2}, \ldots , s_{n} \right\}$ with a neighborhood system $\eta_{i}$. The type-I observed Fisher information regarding the spatial dependence parameter $\beta$ for a given global configuration $\mathbf{X}^{(t)} = \left\{ x_{1}^{(t)}, x_{2}^{(t)}, \ldots , x_{n}^{(t)} \right\}$ is defined as:
\end{mydef}

\begin{equation}
	\phi_{\beta} = \left[ \frac{\partial}{\partial\beta} log~L\left( \vec{\theta}; \mathbf{X}^{(t)} \right)	\right]^{2}
\label{eq:phi_obs_global} 
\end{equation}
\vspace{0.2cm}

An unbiased estimator for the quantity $\phi_{\beta}$ can be obtained by invoking the law of large numbers and approximating equation \eqref{eq:phi_obs_global} by a sample average of the type-I local observed Fisher information $\phi_{\beta}(x_{i})$ along the field:

\begin{equation}
	\hat{\phi}_{\beta} = \frac{1}{n}\sum_{i=1}^{n}\phi_{\beta}(x_{i}) = \frac{1}{n}\sum_{i=1}^{n} \left[ \frac{\partial}{\partial\beta} log~p(x_{i}|{\eta_{i}}, \vec{\theta}) \right]^{2}
\label{eq:phi_obs_global_est} 
\end{equation}
\vspace{0.2cm}

Replacing equation \eqref{eq:phi_local_GMRF} in \eqref{eq:phi_obs_global_est}, we have an expression to compute the type-I observed Fisher information for a global configuration $\mathbf{X}^{(t)}$ modeled by an isotropic pairwise GMRF:  

\begin{equation}
	\hat{\phi}_{\beta} = \frac{1}{n\sigma^{4}}\sum_{i=1}^{n} \left[ \sum_{j \in \eta_{i}}\left(x_{i} - \mu \right) \left(x_{j} - \mu \right) - \beta \sum_{j \in \eta_{i}}\sum_{k \in \eta_{i}}\left(x_{j} - \mu \right) \left(x_{k} - \mu \right)  \right]^{2}
	\label{eq:phi_obs}
\end{equation}
\normalsize

\begin{mydef}
Consider a MRF defined on a lattice $S=\left\{ s_{1}, s_{2}, \ldots , s_{n} \right\}$ with a neighborhood system $\eta_{i}$. The type-II observed Fisher information regarding the spatial dependence parameter $\beta$ for a given global configuration $\mathbf{X}^{(t)} = \left\{ x_{1}^{(t)}, x_{2}^{(t)}, \ldots , x_{n}^{(t)} \right\}$ is defined as:
\end{mydef}

\begin{equation}
	\psi_{\beta} = -\frac{\partial^{2}}{\partial\beta^{2}} log~L\left( \vec{\theta}; \mathbf{X}^{(t)} \right)
\label{eq:psi_obs_global} 
\end{equation}
\vspace{0.2cm}

Similarly to the previous situation, a reasonable approximation for $\psi_{\beta}$ is obtained by taking the sample average of of the type-II local observed Fisher information $\psi_{\beta}(x_{i})$ along the field:

\begin{equation}
	\hat{\psi}_{\beta} = \frac{1}{n}\sum_{i=1}^{n}\psi_{\beta}(x_{i}) = - \frac{1}{n}\sum_{i=1}^{n} \frac{\partial^{2}}{\partial\beta^{2}} log~p(x_{i}|{\eta_{i}}, \vec{\theta}) 
\label{eq:psi_obs_global_est} 
\end{equation}
\vspace{0.2cm}

Replacing equation \eqref{eq:psi_local_GMRF} in \eqref{eq:psi_obs_global_est}, we have an expression to compute the type-II observed Fisher information for a global configuration $\mathbf{X}^{(t)}$ modeled by an isotropic pairwise GMRF:

\begin{equation}
	\hat{\psi}_{\beta} = \frac{1}{n\sigma^{2}}\sum_{i=1}^{n} \sum_{j \in \eta_{i}} \sum_{k \in \eta_{i}} \left( x_{j} - \mu \right) \left( x_{k} - \mu \right)
	\label{eq:psi_obs}
\end{equation}
\vspace{0.2cm}

Therefore, we have two local measures, $\phi_{\beta}(x_{i})$ and $\psi_{\beta}(x_{i})$ that can be assigned to every element of a system modeled by an isotropic pairwise GMRF. Besides, two other global mesures, $\hat{\phi}_{\beta}$ and $\hat{\psi}_{\beta}$, provide the same information but in a larger scale. In the following, we will discuss some interpretations for what is really being measured with the proposed tools.

\subsection{The Role of Fisher information in GMRF models}

At this point, a relevant issue is the interpretation of these Fisher information measures in a complex system modeled by an isotropic pairwise GMRF. Roughly speaking, $\phi_{\beta}(x_{i})$ is the quadratic rate of change of the logarithm of the local likelihood function at $x_{i}$, given a global value of $\beta$. As this global value of $\beta$ determines what would be the expected global behavior (if $\beta$ is large, it is expected a high degree of correlation among the observations and if $\beta$ is close to zero the observations are independent), it is reasonable to admit that configuration patterns showing values of $\phi_{\beta}(x_{i})$ close to zero are more likely to be observed throughout the field, once their likelihood values are high (close to the maximum local likelihood condition). In other words, these patterns are more ``aligned'' to what is considered to be the expected global behavior and therefore the convey little information about the spatial dependence struture (these samples are not informative once they are expected to exist in a system operating at that particular value of inverse temperature).

Now, let us move on to configuration patterns showing high values of $\phi_{\beta}(x_{i})$. Those samples can be considered landmarks, because they convey a large amount of information about the global spatial dependence structure. Roughly speaking, those points are very informative once they are not expected to exist for that particular value of $\beta$ (which guides the expected global behavior of the system). Therefore, type-I local observed Fisher information minimization in GMRF's can be a useful tool in producing novel configuration patterns that are more likely to exist given that chosen value of inverse temperature. Basically, $\phi_{\beta}( x_{i} )$ tell us how informative a given pattern is for that specific global behavior (represented by a single parameter in an isotropic pairwise GMRF model). In summary, this measure quantifies the degree of agreement between an observation $x_{i}$ and the configuration defined by its neighborhood system for a given $\beta$.

As we will see later in the experiments section, typical informative patterns (those showing high values of $\phi_{\beta}( x_{i} )$) in an organized system are located at the boundaries of the regions defining homogeneous areas (since these boundary samples show an unexpected behavior for large $\beta$, which is: there is no strong agreement between $x_{i}$ and its neighbors).

Let us analyze the type-II local observed Fisher information $\psi_{\beta}( x_{i} )$. Informally speaking, this measure can be interpreted as a curvature measure, that is, how curved is the local likelihood function at $x_{i}$. Thus, patterns showing low values of $\psi_{\beta}( x_{i} )$ tend to have a nearly flat local likelihood function. It means that we are dealing with a pattern that could have been observed for a variety of $\beta$ values (a large set of $\beta$ values have approximatelly the same likelihood). An implication of this fact is that in a system dominated by this kind of patterns (patterns for which $\psi_{\beta}( x_{i} )$ is close to zero), small perturbations may cause a sharp change in $\beta$ (and therefore in the expected global behavior). In other words, these patterns are more susceptible to changes once they do not have a ``stable'' configuration (it raises our uncertainty about the true value of $\beta$). 

On the other hand, if the global configuration is mostly composed by patterns exhibiting large values of $\psi_{\beta}( x_{i} )$, changes on the global structure are unlikely to happen (uncertainty on $\beta$ is sufficiently small). Basically, $\psi_{\beta}( x_{i} )$ measures the degree of agreement or dependence among the observations belonging to the same neighborhood system. If at a given $x_{i}$, the observations belonging to $\eta_{i}$ are totally symmetric around the mean value, $\psi_{\beta}( x_{i} )$ would be zero.  It is reasonable to expect that in this situation as there is no information about the induced spatial dependence struture (it means that there is no contextual information available at this point). Notice that the role of $\psi_{\beta}( x_{i} )$ is not the same of $\phi_{\beta}( x_{i} )$. Actually, these two measures are almost inversely related, since if at $x_{i}$ the value of $\phi_{\beta}( x_{i} )$ is high (it is a landmark or boundary pattern), then it is expected that $\psi_{\beta}( x_{i} )$ be low (in decision boundaries or edges the uncertainty about $\beta$ is higher, causing $\psi_{\beta}( x_{i} )$ to be small). In fact, we will observe this behavior in some computational experiments conducted in future sections of the paper.

It is important to mention that these rather informal arguments define the basis for understanding the meaning of the asymptotic variance of maximum pseudo-likelihood estimators, as we will discuss ahead. In summary, $\psi_{\beta}( x_{i} )$ is a measure of how sure or confidente we are about the local spatial dependence structure (at a given point $x_i$), since a high average curvature is desired for predicting the system's global behavior in a reasonable manner (reducing the uncertainty of $\beta$ estimation).

%%%%%%%%%%%%%%%%%%%%%%%%%%%%%%%%%%%%%%%%%%%%%

\section{Expected Fisher Information}

In order to avoid the use of approximations in the computation of the global Fisher information in an isotropic pairwise GMRF, in this section we provide an exact expression for $\hat{\phi}_{\beta}$ and $\hat{\psi}_{\beta}$ as type-I and type-II expected Fisher information. One advantage of using the expected Fisher information instead of its global observed counterpart is the faster computing time. As we will see, instead of computing a single local measure for each observation $x_{i} \in \mathbf{X}$ and then take the average, both $\Phi_{\beta}$ and $\Psi_{\beta}$ expressions depend only on the covariance matrix of the configuration patterns observed along the random field.

\subsection{The Type-I Expected Fisher Information} 

Recall that the type-I expected Fisher information, from now on denoted by $\Phi_{\beta}$, is given by:

\begin{equation}
	\Phi_{\beta} = E\left[ \left( \frac{\partial}{\partial \beta} log~L\left(\vec{\theta}; \mathbf{X}^{(t)} \right) \right)^{2} \right]
	\label{eq:Phi}
\end{equation}
%\vspace{0.2cm}

The type-II expected Fisher information, from now on denoted by $\Psi_{\beta}$, is given by:

\begin{equation}
	\Psi_{\beta} = -E\left[ \frac{\partial^{2}}{\partial \beta^{2}} log~L\left(\vec{\theta}; \mathbf{X}^{(t)} \right) \right]
	\label{eq:Psi}
\end{equation}
%\vspace{0.2cm}

We first proceed to the definition of $\Phi_{\beta}$. Pluging equation \eqref{eq:GMRF_PL} in \eqref{eq:Phi} and after some algebra, we obtain the following expression, which is composed by four main terms:

\begin{align}
	\Phi_{\beta} & = \frac{1}{\sigma^{4}}E\left\{\left[\sum_{s=1}^{n} \left( x_{s} - \mu - \beta\sum_{j \in \eta_s}\left( x_{j} - \mu \right) \right) \left( \sum_{j \in \eta_s}\left( x_{j} - \mu \right) \right) \right]^{2} \right\} = \label{eq:Phi_01} \\ \nonumber \\ \nonumber
	& = \frac{1}{\sigma^{4}}E\left\{ \sum_{s=1}^{n}\sum_{r=1}^{n} \left[ x_{s} - \mu - \beta\sum_{j \in \eta_s}\left( x_{j} - \mu \right) \right]  \left[ x_{r} - \mu - \beta\sum_{k \in \eta_r}\left( x_{k} - \mu \right) \right]  \times \right.  \\ \nonumber 
	& \left. \qquad \qquad \qquad \qquad \left[ \sum_{j \in \eta_s}\left( x_{j} - \mu \right) \right] \left[ \sum_{k \in \eta_r}\left( x_{k} - \mu \right) \right] \right\} = \\ \nonumber \\ \nonumber	
	& = \frac{1}{\sigma^{4}} E\left\{ \sum_{s=1}^{n}\sum_{r=1}^{n} \left[ \left(x_{s} - \mu \right)\left(x_{r} - \mu \right) - \beta\sum_{k \in \eta_r}\left(x_{s} - \mu \right)\left(x_{k} - \mu \right) - \beta\sum_{j \in \eta_s}\left(x_{r} - \mu \right)\left(x_{j} -\mu \right)  \right. \right. \\ \nonumber 
	& \left. \left.  \qquad \qquad \qquad \qquad + \beta^{2}\sum_{j\in\eta_s}\sum_{k\in\eta_r}\left( x_{j} - \mu \right)\left( x_{k} - \mu \right)\right] \left[ \sum_{j\in\eta_s}\sum_{k\in\eta_r}\left( x_{j} - \mu \right)\left( x_{k} - \mu \right) \right]  \right\} \\ \nonumber \\ \nonumber
	& = \frac{1}{\sigma^{4}}\sum_{s=1}^{n}\sum_{r=1}^{n}\left\{ \sum_{j\in\eta_s}\sum_{k\in\eta_r}E\left[ \left( x_{s} - \mu \right)\left( x_{r} - \mu \right)\left( x_{j} - \mu \right)\left( x_{k} - \mu \right) \right]  \right. \\ \nonumber 
	& \left. \qquad\qquad\qquad -\beta\sum_{j\in\eta_s}\sum_{k\in\eta_r}\sum_{l\in\eta_r}E\left[ \left( x_{s} - \mu \right)\left( x_{j} - \mu \right)\left( x_{k} - \mu \right)\left( x_{l} - \mu \right) \right] \right. \\ \nonumber
	&  \left. \qquad\qquad\qquad -\beta\sum_{m\in\eta_s}\sum_{j\in\eta_s}\sum_{k\in\eta_r}E\left[ \left( x_{r} - \mu \right)\left( x_{m} - \mu \right)\left( x_{j} - \mu \right)\left( x_{k} - \mu \right) \right] \right. \\ \nonumber
	& \left.  \qquad\qquad\qquad + \beta^{2}\sum_{m\in\eta_s}\sum_{j\in\eta_s}\sum_{k\in\eta_r}\sum_{l\in\eta_r}E\left[ \left( x_{m} - \mu \right)\left( x_{j} - \mu \right)\left( x_{k} - \mu \right)\left( x_{l} - \mu \right) \right] \right\}
\end{align}

%& = \frac{1}{\sigma^{4}}E\left\{ \sum_{s=1}^{n}\sum_{r=1}^{n} \left[ \left(x_{s} - \mu \right)\left(x_{r} - \mu \right) - \beta\sum_{j \in \eta_r}\left(x_{s} - \mu \right)\left(x_{j} -\mu \right) - \beta\sum_{j \in \eta_s}\left(x_{r} - \mu \right)\left(x_{j} -\mu \right) \right.  \\ \nonumber
%	& \left. \qquad \qquad \qquad \qquad + \beta^{2}\sum_{j\in\eta_s}\sum_{k\in\eta_r}\left( x_{j} - \mu \right)\left( x_{k} - \mu \right)\left] \left[ \sum_{j\in\eta_s}\sum_{k\in\eta_r}\left( x_{j} - \mu \right)\left( x_{k} - \mu \right) \right] \right\}

%\vspace{0.2cm}

%\frac{1}{\sigma^{4}}\sum_{i=1}^{n}~E\left\{ \left[ \left( x_{i} - \mu \right)\sum_{j \in \eta_i}\left( x_{j} - \mu \right) - \beta \left( \sum_{j \in \eta_i}\left( x_{j} - \mu \right) \right)^2 \right]^{2} \right\} = \\ \nonumber
%	 & = \frac{1}{\sigma^{4}}\sum_{i=1}^{n}~E\left\{ \left[ \sum_{j \in \eta_i}\left( x_{i} - \mu \right)\left( x_{j} - \mu \right) \right]^2 - 2\beta\left[\sum_{j \in \eta_i} \left( x_{i} - \mu \right)\left( x_{j} - \mu \right) \right] \left[ \sum_{j \in \eta_i}\left( x_{j} - \mu \right) \right]^2 \right. \\ \nonumber & \left. + \beta^2 \left[ \sum_{j \in \eta_i}\left( x_{j} - \mu \right) \right]^4 \right\} 

Hence, the expression for $\Phi_{\beta}$ is composed by four main terms, each one of them involving a summation of higher-order cross moments. According to the Isserlis' theorem \cite{isserlis1918}, for normally distributed random variables, we can compute higher order moments in terms of the covariance matrix through the following identity:

\begin{align}
	E\left[ X_{1}X_{2}X_{3}X_{4} \right] & = E\left[ X_{1}X_{2} \right]E\left[ X_{3}X_{4} \right] + E\left[ X_{1}X_{3} \right]E\left[ X_{2}X_{4} \right] + E\left[ X_{2}X_{3} \right]E\left[ X_{1}X_{4} \right]
\end{align}

Then, the first term of \eqref{eq:Phi_01} is reduced to:

%Expanding the first term of the previous expression gives us the following expression:
%
%\begin{align}
%	E\left\{ \left[ \sum_{j \in \eta_i}\left( x_{i} - \mu \right)\left( x_{j} - \mu \right) \right]^2\right\} & = 
%	E\left[ \sum_{j \in \eta_i}\sum_{k \in \eta_i}\left( x_{i} - \mu \right)^2 \left( x_{j} - \mu \right) \left( x_{k} - \mu \right) \right] = \label{eq:Phi_Termo1} \\ \nonumber \\ \nonumber
%	& = \sum_{j \in \eta_i}\sum_{k \in \eta_i}E\left[ \left( x_{i} - \mu \right) \left( x_{i} - \mu \right) \left( x_{j} - \mu \right) \left( x_{k} - \mu \right) \right] 	
%\end{align}
%\vspace{0.2cm}

%\vspace{0.2cm}

\begin{align}
 & \sum_{j\in\eta_s}\sum_{k\in\eta_r}E\left[ \left( x_{s} - \mu \right)\left( x_{r} - \mu \right)\left( x_{j} - \mu \right)\left( x_{k} - \mu \right) \right] = \label{eq:Phi_Termo1b} \\ \nonumber \\ \nonumber
 & \sum_{j \in \eta_s}\sum_{k \in \eta_r} \left\{ E\left[ \left( x_{s} - \mu \right)\left( x_{r} - \mu \right) \right] E\left[ \left( x_{j} - \mu \right)\left( x_{k} - \mu \right) \right]\right. \\ \nonumber & + \left. E\left[ \left( x_{s} - \mu \right)\left( x_{j} - \mu \right) \right]E\left[ \left( x_{r} - \mu \right) \left( x_{k} - \mu \right) \right] \right.  \\ \nonumber & + \left. E\left[ \left( x_{r} - \mu \right)\left( x_{j} - \mu \right) \right]E\left[ \left( x_{s} - \mu \right) \left( x_{k} - \mu \right) \right] 	\right\} = \\ \nonumber \\ \nonumber
 & \sum_{j \in \eta_s}\sum_{k \in \eta_r} \left[ \sigma_{sr} \sigma_{jk} + \sigma_{sj} \sigma_{rk} + \sigma_{rj} \sigma_{sk} \right]
\end{align} where $\sigma_{sr}$ denotes the covariance between variables $x_{s}$ and $x_{r}$. (note that in a MRF we have $\sigma_{sr} = 0$ if $x_{r} \notin \eta_{s}$). We now proceed to the expansion of the second main term of \eqref{eq:Phi_01}. Similarly, by applying the Isserslis' identity we have:

\begin{align}
	& \sum_{j\in\eta_s}\sum_{k\in\eta_r}\sum_{l\in\eta_r}E\left[ \left( x_{s} - \mu \right)\left( x_{j} - \mu \right)\left( x_{k} - \mu \right)\left( x_{l} - \mu \right) \right] = \sum_{j\in\eta_s}\sum_{k\in\eta_r}\sum_{l\in\eta_r}\left[ \sigma_{sj}\sigma_{kl} + \sigma_{sk}\sigma_{jl} + \sigma_{jk}\sigma_{sl} \right]
	\label{eq:Phi_Termo2}
\end{align}
%\vspace{0.2cm}

The thrid term of \eqref{eq:Phi_01} can be rewritten as:

\begin{align}
	& \sum_{m\in\eta_s}\sum_{j\in\eta_s}\sum_{k\in\eta_r}E\left[ \left( x_{r} - \mu \right)\left( x_{m} - \mu \right)\left( x_{j} - \mu \right)\left( x_{k} - \mu \right) \right] = \label{eq:Phi_Termo3} \\ \nonumber & = \sum_{m\in\eta_s}\sum_{j\in\eta_s}\sum_{k\in\eta_r}\left[ \sigma_{rm}\sigma_{jk} + \sigma_{rj}\sigma_{mk} + \sigma_{mj}\sigma_{rk} \right]
\end{align}
\vspace{0.25cm}

Finally, the fourth term of is:

\begin{align}
	& \sum_{m\in\eta_s}\sum_{j\in\eta_s}\sum_{k\in\eta_r}\sum_{l\in\eta_r}E\left[ \left( x_{m} - \mu \right)\left( x_{j} - \mu \right)\left( x_{k} - \mu \right)\left( x_{l} - \mu \right) \right] = \label{eq:Phi_Termo4} \\ \nonumber & = \sum_{m\in\eta_s}\sum_{j\in\eta_s}\sum_{k\in\eta_r}\sum_{l\in\eta_r}\left[ \sigma_{mj}\sigma_{kl} + \sigma_{mk}\sigma_{jl} + \sigma_{ml}\sigma_{jk} \right]	
\end{align}
\vspace{0.25cm}

Therefore, by combining expressions \eqref{eq:Phi_Termo1b}, \eqref{eq:Phi_Termo2}, \eqref{eq:Phi_Termo3} and \eqref{eq:Phi_Termo4} we have the complete expression for $\Phi_{\beta}$, the type-I expected Fisher information for an isotropic pairwise GMRF model regarding the inverse temperature parameter, as:

\begin{align}
	\Phi_{\beta} = \frac{1}{\sigma^{4}}\sum_{s=1}^{n}\sum_{r=1}^{n} & \left\{ \sum_{j \in \eta_s}\sum_{k \in \eta_r} \left[ \sigma_{sr} \sigma_{jk} + \sigma_{sj} \sigma_{rk} + \sigma_{rj} \sigma_{sk} \right] \right. \label{eq:Phi_Completa}  \\ \nonumber & \left. - \beta\sum_{j\in\eta_s}\sum_{k\in\eta_r}\sum_{l\in\eta_r}\left[ \sigma_{sj}\sigma_{kl} + \sigma_{sk}\sigma_{jl} + \sigma_{jk}\sigma_{sl} \right] \right. \\ \nonumber & \left. - \beta\sum_{m\in\eta_s}\sum_{j\in\eta_s}\sum_{k\in\eta_r}\left[ \sigma_{rm}\sigma_{jk} + \sigma_{rj}\sigma_{mk} + \sigma_{mj}\sigma_{rk} \right] \right. \\ \nonumber & \left.  + \beta^{2}\sum_{m\in\eta_s}\sum_{j\in\eta_s}\sum_{k\in\eta_r}\sum_{l\in\eta_r}\left[ \sigma_{mj}\sigma_{kl} + \sigma_{mk}\sigma_{jl} + \sigma_{ml}\sigma_{jk} \right] \right\}
\end{align}
\vspace{0.25cm}

However, since we are interested in studying how the spatial correlations change as the system evolves, we need to estimate a value for $\Phi_{\beta}$ given a single global state $\mathbf{X}^{(t)} = \left\{ x_{1}^{(t)}, x_{2}^{(t)}, \ldots , x_{n}^{(t)} \right\}$. Hence, to compute $\Phi_{\beta}$ from a single static configuration $\mathbf{X}^{(t)}$ (a photograph of the system at a given moment), we consider $n = 1$ in the previous equation, which means, among other things, that $s = r$ (which implies $\eta_s = \eta_r$) and observations belonging to different neighborhoods are independent from each other (since we are dealing with a pairwise interaction Markovian process). 

Before proceeding, we would like to clarify some points regarding the estimation of the $\beta$ parameter and the computation of the expected Fisher information in the isotropic pairwise GMRF model. Basically, there are two main possibilities: 1) the parameter is spatially-invariant, which means that we have a unique value $\hat{\beta}^{(t)}$ for a global configuration of the system $\mathbf{X}^{(t)}$ (this is our assumption); or 2) the parameter is spatially-variant, which means that we have a set of $\hat{\beta}_{s}$ values, for $s=1,2,\ldots,n$, each one of them estimated from $\mathbf{X}_{s} = \left\{ x_{s}^{(1)}, x_{s}^{(2)}, \ldots, x_{s}^{(t)} \right\}$ (we are observing the outcomes of a random pattern along time in a fixed position of the lattice). When we are dealing with the first model ($\beta$ is spattialy-invariant), all possible observation patters (samples) are extracted from the global configuration by a sliding window (with the shape of the neighborhood system) that moves through the lattice at a fixed time instant $t$. In this case, we are interested in studying the spatial correlations, not the temporal ones. In other words, we would like to investigate how the the spatial structure of a GMRF model is related to Fisher information (this is exactly the scenario described above, for which $n = 1$). Our motivation here is to characterize, via information-theoretic measures, the behavior of the system as it evolves from states of minimum entropy to states of maximum entropy (and vice versa) by providing a geometrical tool based on the definition of the \emph{Fisher curve}, which will be introduced in the following sections.

Therefore, in our case ($n = 1$), equation \eqref{eq:Phi_Completa} is simplified to (unifiyng $s = r = i$ to express the covariances between the random variables in the neighborhood system):

% Equacao final simplificada quando n = 1 e por consequencia i=j
\begin{equation}
	\Phi_{\beta} = \frac{1}{\sigma^4}\left\{ \sum_{j \in \eta_i}\sum_{k \in \eta_i} \left[ \sigma^2\sigma_{jk} + 2\sigma_{ij}\sigma_{ik} \right] - 2\beta\sum_{j \in \eta_i}\sum_{k \in \eta_i}\sum_{l \in \eta_i}\left[ \sigma_{ij}\sigma_{kl} + \sigma_{ik}\sigma_{jl} + \sigma_{il}\sigma_{jk} \right] \right.  
\label{eq:Phi_OK} 	
\end{equation}	
\begin{equation}	
	 \left. + \beta^{2}\sum_{j \in \eta_i}\sum_{k \in \eta_i}\sum_{l \in \eta_i}\sum_{m \in \eta_i}\left[ \sigma_{jk}\sigma_{lm} + \sigma_{jl}\sigma_{km} + \sigma_{jm}\sigma_{kl} \right]  \right\} \nonumber
\end{equation}

%Note that, in this case, we have $n = 1$ in the summation of the expected Fisher information ($\Phi_{\beta}$) since we want to compute this value for a static global configuration of the system.

\subsection{The Type-II Expected Fisher Information}

Following the same methodology of replacing the likelihood function by the pseudo-likelihood function of the GMRF model, a closed form expression for $\Psi_{\beta}$ is developed. Pluging equation \eqref{eq:GMRF_PL} into \eqref{eq:Psi} leads us to:

\begin{align}
	& \Psi_{\beta} = \frac{1}{\sigma^{2}}\sum_{i=1}^{n}~E\left\{ \left[ \sum_{x_j \in \eta_i}\left( x_{j} - \mu \right) \right]^2 \right\} \\ \nonumber \\ \nonumber
	& = \frac{1}{\sigma^{2}}\sum_{i=1}^{n}~E\left[ \sum_{x_j \in \eta_i}\sum_{x_k \in \eta_i} \left( x_{j} - \mu \right)\left( x_{k} - \mu \right) \right] = \\ \nonumber \\ \nonumber
	& = \frac{1}{\sigma^{2}}\sum_{i=1}^{n}\left\{ \sum_{x_j \in \eta_i}\sum_{x_k \in \eta_i}~E\left[ \left( x_{j} - \mu \right) \left( x_{k} - \mu \right) \right] \right\} 
	= \frac{1}{\sigma^2} \sum_{i=1}^{n} \sum_{j \in \eta_i}\sum_{k \in \eta_i} \sigma_{jk} \label{eq:Psi_OK}
\end{align}
\vspace{0.2cm}

Note that unlike $\Phi_{\beta}$, $\Psi_{\beta}$ does not depend explicity on $\beta$ (inverse temperature). As we have seen before, $\Phi_{\beta}$ is a quadratic function of the spatial dependence parameter.

In order to simplify the notations and also to make computations easier, the expressions for $\Phi_{\beta}$ and $\Psi_{\beta}$ can be rewritten in a matrix-vector form. Let $\Sigma_{p}$ be the covariance matrix of the random vectors $\vec{p}_{i}, i = 1,2,\ldots,n$, obtained by lexicographic ordering the local configuration patterns $x_{i} \cup \eta_{i}$. Thus, considering a neighborhood system $\eta_{i}$ of size $K$, we have $\Sigma_{p}$ given by a $(K + 1) \times (K + 1)$ symmetric matrix (for $K+1$ odd, i.e., $K = 4, 8, 12, \ldots$):

\[
 \Sigma_{p} =
 \begin{pmatrix}
  \sigma_{1,1} & \cdots & \sigma_{1,K/2} & \sigma_{1,(K/2)+1} & \sigma_{1,(K/2)+2} & \cdots & \sigma_{1,K+1} \\
  \vdots  & \vdots & \vdots & \vdots & \vdots & \vdots & \vdots  \\
  \sigma_{K/2,1} & \cdots & \sigma_{K/2,K/2} & \sigma_{K/2,(K/2)+1} & \sigma_{K/2,(K/2)+2} & \cdots & \sigma_{K/2,K+1} \\
  \sigma_{(K/2)+1,1} & \cdots & \sigma_{(K/2)+1,K/2} & \sigma_{(K/2)+1,(K/2)+1} & \sigma_{(K/2)+1,(K/2)+2} & \cdots & \sigma_{(K/2)+1,K+1} \\
  \sigma_{(K/2)+2,1} & \cdots & \sigma_{(K/2)+2,K/2} & \sigma_{(K/2)+2,(K/2)+1} & \sigma_{(K/2)+2,(K/2)+2} & \cdots & \sigma_{(K/2)+2,K+1} \\
  \vdots   & \vdots & \vdots & \vdots & \vdots & \vdots & \vdots  \\
  \sigma_{K+1,1} & \cdots & \sigma_{K+1,K/2} & \sigma_{K+1,(K/2)+1} & \sigma_{K+1,(K/2)+2} & \cdots & \sigma_{K+1,K+1}
 \end{pmatrix}
 \label{eq:SigmaP} 	
\]
\vspace{0.2cm}

Let $\Sigma_{p}^{-}$ be the submatrix of dimensions $K \times K$ obtained by removing the central row and central column of $\Sigma_{p}$ (the covariances between $x_{i}$ and each one of its neighbors $x_{j}$). Then for $K+1$ odd, we have:

\begin{equation}
 \Sigma_{p}^{-} =
 \begin{pmatrix}
  \sigma_{1,1} & \cdots & \sigma_{1,K/2} & \sigma_{1,(K/2)+2} & \cdots & \sigma_{1,K+1} \\
  \vdots  & \vdots & \vdots & \vdots & \vdots & \vdots  \\
  \sigma_{K/2,1} & \cdots & \sigma_{K/2,K/2} & \sigma_{K/2,(K/2)+2} & \cdots & \sigma_{K/2,K+1} \\
  \sigma_{(K/2)+2,1} & \cdots & \sigma_{(K/2)+2,K/2} & \sigma_{(K/2)+2,(K/2)+2} & \cdots & \sigma_{(K/2)+2,K+1} \\
  \vdots & \vdots & \vdots & \vdots & \vdots & \vdots  \\
  \sigma_{K+1,1} & \cdots & \sigma_{K+1,K/2} & \sigma_{K+1,(K/2)+2} & \cdots & \sigma_{K+1,K+1}
 \end{pmatrix}
 \label{eq:SigmaP_minus} 	
\end{equation}
\vspace{0.2cm}

 Thus, $\Sigma_{p}^{-}$ is a matrix that stores only the covariances among the neighboring variables. Also, let $\vec{\rho}$ be the vector of dimensions $K \times 1$ formed by all the elements of the central row of $\Sigma_{p}$, excluding the middle one (which is a variance actually), that is:

\begin{equation}
	\vec{\rho} = \left[\sigma_{(K/2)+1,1} ~~~ \cdots ~~~ \sigma_{(K/2)+1,K/2} ~~~ \sigma_{(K/2)+1,(K/2)+2} ~~~ \cdots ~~~  \sigma_{(K/2)+1,K+1} \right]
	\label{eq:rho}
\end{equation}
\vspace{0.2cm}

   Therefore, we can rewrite equation \eqref{eq:Phi_OK} (for $n = 1$) using Kronecker products. The following definition provides a fast way to compute $\Phi_{\beta}$ exploring these tensor products.
   
\begin{mydef}
Let an isotropic pairwise GMRF be defined on a lattice $S=\left\{ s_{1}, s_{2}, \ldots , s_{n} \right\}$ with a neighborhood system $\eta_{i}$ of size $K$ (usual choices for $K$ are even values: 4, 8, 12, 20 or 24). Assuming that $\mathbf{X^{(t)}}=\{x_{1}^{(t)}, x_{2}^{(t)}, \ldots, x_{n}^{(t)} \}$ denotes the global configuration of the system at time $t$, and $\vec{\rho}$ and $\Sigma_{p}^{-}$ are defined as equations \eqref{eq:rho} and \eqref{eq:SigmaP_minus}, the type-I expected Fisher information $\Phi_{\beta}$ for this state $\mathbf{X^{(t)}}$ is:
\end{mydef}   
   
\begin{align}
	\Phi_{\beta} = \frac{1}{\sigma^4}\left[ \sigma^2 \left\|\Sigma_{p}^{-}\right\|_{+} + 2 \left\|\vec{\rho}\otimes\vec{\rho}^T\right\|_{+} - 6\beta \left\| \vec{\rho}^{T} \otimes \Sigma_{p}^{-} \right\|_{+} + 3\beta^2 \left\| \Sigma_{p}^{-} \otimes \Sigma_{p}^{-} \right\|_{+} \right]  
\label{eq:Phi_OK_Kron}
\end{align} where $\left\| A \right\|_{+}$ denotes the summation of all the entries of the matrix $A$ (not to be confused with a matrix norm) and $\otimes$ denotes the Kronecker (tensor) product. From an information geometry perspective, the presence of tensor products indicates the intrinsic differential geometry of a manifold in the form of the Riemman curvature tensor \cite{Amari}. Note that all the necessary information for computing the Fisher information is somehow encoded in the covariance matrix of the local configuration patterns, $(x_{i} \cup \eta_{i}), i = 1,2,\ldots,n$, as it would be expected in case of Gaussian variables (second-order statistics). The same procedure is applied to the type-II expected Fisher information.

\vspace{0.2cm}

\begin{mydef}
Let an isotropic pairwise GMRF be defined on a lattice $S=\left\{ s_{1}, s_{2}, \ldots , s_{n} \right\}$ with a neighborhood system $\eta_{i}$ of size $K$ (usual choices for $K$ are 4, 8, 12, 20 or 24). Assuming that $\mathbf{X^{(t)}}=\{x_{1}^{(t)}, x_{2}^{(t)}, \ldots, x_{n}^{(t)} \}$ denotes the global configuration of the system at time $t$ and $\Sigma_{p}^{-}$ is defined as equation \eqref{eq:SigmaP_minus}, the type-II expected Fisher information $\Psi_{\beta}$ for this state $\mathbf{X^{(t)}}$ is given by:
\end{mydef}

\begin{align}
	\Psi_{\beta} = \frac{1}{\sigma^2}\left\| \Sigma_{p}^{-} \right\|_{+}
	\label{eq:Psi_OK_Kron}
\end{align} 

\subsection{Information Equilibrium in GMRF models}

From the definition of both $\Phi_{\beta}$ and $\Psi_{\beta}$, a natural question that raises would be: under what conditions do we have $\Phi_{\beta} = \Psi_{\beta}$ in an isotropic pairwise GMRF model? As we can see from equations \eqref{eq:Phi_OK_Kron} and \eqref{eq:Psi_OK_Kron}, the difference between $\Phi_{\beta}$ and $\Psi_{\beta}$, from now on denote by $\Delta_{\beta}\left(\vec{\rho}, \Sigma_{p}^{-} \right)$ is simply:

\begin{align}
	\Delta_{\beta} \left( \vec{\rho}, \Sigma_{p}^{-}\right) = \frac{1}{\sigma^{4}}\left( 2 \left\|\vec{\rho}\otimes\vec{\rho}^T\right\|_{+} - 6\beta \left\| \vec{\rho}^{T} \otimes \Sigma_{p}^{-} \right\|_{+} + 3\beta^2 \left\| \Sigma_{p}^{-} \otimes \Sigma_{p}^{-} \right\|_{+} \right)
	\label{eq:Equilibrium} 
\end{align} 

Then, intuitively, the condition for information equality is achieved when $\Delta_{\beta}\left(\vec{\rho}, \Sigma_{p}^{-} \right) = 0$. As $\Delta_{\beta}\left(\vec{\rho}, \Sigma_{p}^{-} \right)$ is a simple quadratic function of the inverse temperature parameter $\beta$, we can easily find that the value $\beta^{*}$ for which $\Delta_{\beta}\left(\vec{\rho}, \Sigma_{p}^{-} \right) = 0$ is:

\begin{align}
	\beta^{*} = \frac{\left\| \vec{\rho}^{T} \otimes \Sigma_{p}^{-} \right\|_{+}}{\left\| \Sigma_{p}^{-} \otimes \Sigma_{p}^{-} \right\|_{+}} \pm \frac{\sqrt{3}}{3}\frac{\sqrt{3 \left\| \vec{\rho}^{T} \otimes \Sigma_{p}^{-} \right\|_{+}^{2} - 2 \left\| \Sigma_{p}^{-} \otimes \Sigma_{p}^{-} \right\|_{+} \left\|\vec{\rho}\otimes\vec{\rho}^T\right\|_{+}}}{\left\| \Sigma_{p}^{-} \otimes \Sigma_{p}^{-} \right\|_{+}} \label{eq:solution} 
\end{align} provided that $3\left\| \vec{\rho}^{T} \otimes \Sigma_{p}^{-} \right\|_{+}^{2} \geq 2 \left\| \Sigma_{p}^{-} \otimes \Sigma_{p}^{-} \right\|_{+} \left\|\vec{\rho}\otimes\vec{\rho}^T\right\|_{+}$ and $\left\| \Sigma_{p}^{-} \otimes \Sigma_{p}^{-} \right\|_{+} \neq 0$. Note that if $\left\|\vec{\rho}\otimes\vec{\rho}^T\right\|_{+} = 0$, then one solution for the above equation is $\beta^{*} = 0$. In other words, when $\sigma_{ij} = 0, \forall j \in \eta_{i}$ (no correlation between $x_{i}$ and its neighbors $x_{j}$), information equilibrium is achieved for $\beta^{*} = 0$, which in this case is the maximum pseudo-likelihood estimative of $\beta$, since in this matrix-vector notation $\hat{\beta}_{MPL}$ is given by:

\begin{equation}
	\hat{\beta}_{MPL} = \frac{\displaystyle\sum_{j \in \eta_i}\hat{\sigma}_{ij}}{\displaystyle\sum_{j \in \eta_i}\displaystyle\sum_{k \in \eta_i}\hat{\sigma}_{jk}} = \frac{\left\| \vec{\rho} \right\|_{+}}{\left\| \Sigma_{p}^{-} \right\|_{+}}
\end{equation}

In the isotropic pairwise GMRF model, if $\beta = 0$ them we have $\left\| \vec{\rho} \right\|_{+} = 0$ and as a consequence $\Phi_{\beta} = \Psi_{\beta}$. However, the opposite is not necessarily true, that is, we may observe that $\Phi_{\beta} = \Psi_{\beta}$ for a non-zero $\beta$. One example is for $\beta^{*}$, a solution of $\Delta_{\beta}\left(\vec{\rho}, \Sigma_{p}^{-} \right) = 0$. 

%%%%%%%%%%%%%%%%%%%%%%%%%%%%%%%%%%%%%%%%%%%%%

\section{Entropy in Isotropic Pairwise GMRF's}

Our definition of entropy is done by repeating the same process employed to derive $\Phi_{\beta}$ and $\Psi_{\beta}$. Knowing that the entropy of random variable $x$ is defined by the expected value of self-information, given by $-log~p(x)$, it can be thought as a probability-based counterpart to the Fisher information.

\begin{mydef}
Let an isotropic pairwise GMRF be defined on a lattice $S=\left\{ s_{1}, s_{2}, \ldots , s_{n} \right\}$ with a neighborhood system $\eta_{i}$. Assuming that $\mathbf{X^{(t)}}=\{x_{1}^{(t)}, x_{2}^{(t)}, \ldots, x_{n}^{(t)} \}$ denotes the global configuration of the system at time $t$, then the entropy $H_{\beta}$ for this state $\mathbf{X^{(t)}}$ is given by:
\end{mydef}

\begin{align}
	& H_{\beta} = - E\left[ log~L\left(\vec{\theta}; \mathbf{X}^{(t)} \right) \right] = - E\left[ log \prod_{i=1}^{n}p\left( x_{i}|\eta_{i}, \vec{\theta} \right) \right] \label{eq:entropia1} = \\ \nonumber \\ \nonumber & = \frac{n}{2}log\left( 2\pi\sigma^2 \right) + \frac{1}{2\sigma^2}\sum_{i=1}^{n}E\left\{ \left[ x_{i} - \mu - \beta\sum_{j \in \eta_i}\left( x_{j} - \mu \right) \right]^2 \right\} = \\ \nonumber \\ \nonumber & = \frac{n}{2}log\left( 2\pi\sigma^2 \right) + \frac{1}{2\sigma^2}\sum_{i=1}^{n}\left\{ E\left[ \left( x_{i} - \mu \right)^2 \right] -2\beta E\left[ \sum_{j \in \eta_i}\left( x_{i} - \mu \right)\left( x_{j} - \mu \right) \right] \right. \\ \nonumber & + \left. \beta^{2} E\left\{ \left[ \sum_{j \in \eta_i}\left( x_{j} - \mu \right) \right]^2 \right\} \right\} 
\end{align}
\vspace{0.2cm}

After some algebra the expression for $H_{\beta}$ becomes:

\begin{align}
	& H_{\beta} = \frac{n}{2}log\left( 2\pi\sigma^2 \right) + \frac{1}{2\sigma^2}\sum_{i=1}^{n}\left\{\sigma^2 - 2\beta \sum_{j \in \eta_i}\sigma_{ij} + \beta^{2}\sum_{j \in \eta_i}\sum_{k \in \eta_i}\sigma_{jk} \right\} = \\ \nonumber \\ \nonumber & = \Big[ \frac{n}{2}log(2\pi\sigma^{2}) + \frac{n}{2} \Big] - \frac{\beta}{\sigma^{2}}\sum_{i=1}^{n}\left[ \sum_{j \in \eta_{i}}\sigma_{ij} \right] + \frac{\beta^{2}}{2\sigma^{2}}\sum_{i=1}^{n}\left[ \sum_{j \in \eta_{i}} \sum_{k \in \eta_{i}} \sigma_{jk} \right]
%= \\ \\ & = \frac{\beta^{2}}{2} \Psi_{\beta} - n\beta \sum_{j \in \eta_i}\sigma_{ij} + \Big[ \frac{n}{2}log(2\pi) + \frac{n}{2}log(\sigma^2) + \frac{n}{2} \Big] 
\end{align} 
\vspace{0.2cm}

Using the same matrix-vector notation introduced in the previous sections, we can further simplify the expression for $H_{\beta}$ (considering $n = 1$).

\begin{mydef}
Let an isotropic pairwise GMRF be defined on a lattice $S=\left\{ s_{1}, s_{2}, \ldots , s_{n} \right\}$ with a neighborhood system $\eta_{i}$. Assuming that $\mathbf{X^{(t)}}=\{x_{1}^{(t)}, x_{2}^{(t)}, \ldots, x_{n}^{(t)} \}$ denotes the global configuration of the system at time $t$, and $\vec{\rho}$ and $\Sigma_{p}^{-}$ are defined as equations \eqref{eq:rho} and \eqref{eq:SigmaP_minus}, the entropy $H_{\beta}$ for this state $\mathbf{X^{(t)}}$ is given by:
\end{mydef}

\begin{align}     
	H_{\beta} = H_{G} - \left[ \frac{\beta}{\sigma^{2}}\left\| \vec{\rho} \right\|_{+} - \frac{\beta^{2}}{2\sigma^{2}}\left\| \Sigma_{p}^{-}\right\|_{+} \right] = H_{G} - \left[ \frac{\beta}{\sigma^{2}}\left\| \vec{\rho} \right\|_{+} - \frac{\beta^{2}}{2}\Psi_{\beta} \right]
\end{align} where $H_{G}$ denotes the entropy of a Gaussian random variable with variance $\sigma^{2}$ and $\Psi_{\beta}$ is the type-II expected Fisher information.

Note that Shannon entropy is a quadratic function of the spatial dependence parameter $\beta$. Since the coefficient of the quadratic term is strictly non-negative ($\Psi_{\beta}$ is the type-II expected Fisher information), entropy is a convex function of $\beta$. Also, as expected, when $\beta = 0$ and there is no induced spatial dependence in the system, the resulting expression for $H_{\beta}$ is the usual entropy of a Gaussian random variable, $H_{G}$. Thus, there is a value $\hat{\beta_{MH}}$ for the inverse temperature parameter which minimizes the entropy of the system. In fact, $\hat{\beta}_{MH}$ is given by:

\begin{align}   
	\frac{\partial H_{\beta}}{\partial \beta} = & \frac{\beta}{\sigma^{2}}\left\| \Sigma_{p}^{-} \right\|_{+} - \frac{1}{\sigma^{2}}\left\| \vec{\rho} \right\|_{+} = 0 \\ \nonumber \\ \nonumber
	\hat{\beta}_{MH} = & \frac{\left\| \vec{\rho} \right\|_{+}}{\left\| \Sigma_{p}^{-} \right\|_{+}} = \hat{\beta}_{MPL}
\end{align} showing that the maximum pseudo-likelihood and the minimum-entropy estimatives are equivalent in an isotropic pairwise GMRF model. Moreover, using the derived equations we see a relationship between $\Phi_{\beta}, \Psi_{\beta}$ and $H_{\beta}$:

\begin{align}   
	 \Phi_{\beta} - \Psi_{\beta}  =~ & \Delta_{\beta} \left( \vec{\rho}, \Sigma_{p}^{-} \right) \\ \nonumber \\ \nonumber
	 \frac{\partial^{2} H_{\beta}}{\partial \beta^{2}}  =~ & \Psi_{\beta} 
\end{align} where the functional $\Delta_{\beta} \left( \vec{\rho}, \Sigma_{p}^{-} \right)$ that represents the difference between $\Phi_{\beta}$ and $\Psi_{\beta}$ is defined by equation (\ref{eq:Equilibrium}). These equations relate the entropy and one form of Fisher information ($\Psi_{\beta}$) in GMRF models, showing that $\Psi_{\beta}$ can be roughly viewed as the curvature of $H_{\beta}$. In this sense, in a hypothetical information equilibrium condition $\Psi_{\beta} = \Phi_{\beta} = 0$, the entropy's curvature would be null ($H_{\beta}$ would never change). These results suggest that an increase in the value of $\Psi_{\beta}$, which means stability (a measure of agreement between the neighboring observations of a given point), contributes to curve, and therefore to induce a change in the entropy of the system. In this context, the analysis of the Fisher information could bring us insights in predicting the entropy of a system.

\section{Asymptotic Variance of MPL Estimators} \label{sec:var}

It is known from the statistical inference literature, that unbiasedness is a property that is not granted by maximum likelihood estimation neither by maximum pseudo-likelihood (MPL) estimation. Actually, there is no universal method that guarantees the existence of unbiased estimators for a fixed $n$-size sample. Often, in the exponential family of distributions, maximum likelihood estimators (MLE's) coincide with the UMVU (\emph{Uniform Minimum Variance Unbiased}) estimators because MLE's are functions of complete sufficient statistics. There is an impoertant result in statistical inference that shows that if the MLE is unique, then it is a function of sufficient statistics. We could enumerate and make a huge list of several properties that make maximum likelihood estimation a reference method~\cite{Lehmann1983, Bickel1991, Casella2002}.
One of the most important properties concerns the asymptotic behavior of MLE's: when we make the sample size grow infinitely $(n\rightarrow\infty)$, MLE's becomes asymptotically unbiased and efficient. Unfortunately, there is no result showing that the same occurs in maximum pseudo-likelihood estimation. The objective of this section is to propose a closed expression for the asymptotic variance of the maximum pseudo-likelihood of $\beta$ in an isotropic pairwise GMRF model. Unsurprisingly, this variance is completely defined as a function of both forms of expected Fisher information, $\Psi_{\beta}$ and $\Phi_{\beta}$, as for general values of the inverse temperature parameter, the information equality condition fails.

\subsection{The Asymptotic Variance of the Inverse Temperature Parameter}

	In mathematical statistics, asymptotic evaluations uncover several fundamental properties of inference methods, providing a powerful and general tool for studying and characterizing the behavior of estimators. In this Section our objective is to derive an expression for the asymptotic variance of the maximum pseudo-likelihood estimator of the inverse temperature parameter ($\beta$) in isotropic pairwise GMRF models. It is known from the statistical inference literature that both maximum likelihood and maximum pseudo-likelihood estimators share two important properties: consistency and asymptotic normality ~\cite{Jensen, Winkler}. It is possible, therefore, to completely characterize their behaviors in the limiting case. In other words, the asymptotic distribution of $\hat{\beta}_{MPL}$ is normal, centered around the real parameter value (since consistency means that the estimator is asymptotically unbiased), with the asymptotic variance representing the uncertainty about how far we are from the mean (real value). From a statistical perspective, $\hat{\beta}_{MPL} \approx N\left(\beta, \upsilon_{\beta} \right)$, where $\upsilon_{\beta}$ denotes the asymptotic variance of the maximum pseudo-likelihood estimator. It is known that the asymptotic covariance matrix of maximum pseudo-likelihood estimators is given by~\cite{Covariance}:
	
\begin{equation}
C(\vec{\theta}) = H^{-1}(\vec{\theta})J(\vec{\theta})H^{-1}(\vec{\theta})
\label{CovMatrix} 
\end{equation} with

\begin{align}
	H(\vec{\theta}) = E_{\beta}\left[\nabla^{2}log~L\left(\vec{\theta}; \mathbf{X}^{(t)} \right)\right] \\ %\nonumber \\ 
	J(\vec{\theta}) = Var_{\beta}\left[\nabla log~L\left(\vec{\theta}; \mathbf{X}^{(t)} \right)\right] 
\end{align} where $H$ and $J$ denote, respectively, the Jacobian and Hessian matrices regarding the logarithm of the pseudo-likelihood function. Thus, considering the parameter of interest, $\beta$, we have the following definition for its asymptotic variance $\upsilon_{\beta}$ (the derivatives are taken with respect to $\beta$):

\begin{align}
\upsilon_{\beta} = \frac{Var_{\beta}\left[ \frac{\partial}{\partial\beta} log~L\left(\vec{\theta}; \mathbf{X}^{(t)} \right) \right]}{E_{\beta}^{2}\left[ \frac{\partial^{2}}{\partial\beta^{2}} log~L\left(\vec{\theta}; \mathbf{X}^{(t)} \right) \right]} = \frac{E_{\beta}\left[ \left( \frac{\partial}{\partial\beta} log~L\left(\vec{\theta}; \mathbf{X}^{(t)} \right) \right)^{2} \right] - E_{\beta}^{2}\left[ \frac{\partial}{\partial\beta} log~L\left(\vec{\theta}; \mathbf{X}^{(t)} \right) \right]}{E_{\beta}^{2}\left[ \frac{\partial^{2}}{\partial\beta^{2}} log~L\left(\vec{\theta}; \mathbf{X}^{(t)} \right) \right]}
\label{eq:variances}
\end{align} 
\vspace{0.2cm}

However, note that the expected value of the first derivative of $log~L\left(\vec{\theta}; \mathbf{X}^{(t)} \right)$ with relation to $\beta$ is zero:

\begin{align}
	E\left[ \frac{\partial}{\partial\beta}log~L\left(\vec{\theta}; \mathbf{X}^{(t)} \right) \right] = \frac{1}{\sigma^2}\sum_{i=1}^{n}\left\{ E\left[ x_{i} - \mu \right] - \beta \sum_{j \in \eta_i} E\left[ x_{j} - \mu \right] \right\} = 0 
\end{align}

Therefore, the second term of the numerator of \eqref{eq:variances} vanishes and the final expression for the asymptotic variance of the inverse temperature parameter is given as the ratio between $\Phi_{\beta}$ and $\Psi_{\beta}^{2}$:

\begin{align}
	\upsilon_{\beta} & = \frac{1}{\left[\displaystyle\sum_{j \in \eta_i}\displaystyle\sum_{k \in \eta_i}\sigma_{jk}\right]^{2}}\left\{ \displaystyle\sum_{j \in \eta_i}\displaystyle\sum_{k \in \eta_i} \left[ \sigma^2\sigma_{jk} + 2\sigma_{ij}\sigma_{ik} \right] - 2\beta\displaystyle\sum_{j \in \eta_i}\displaystyle\sum_{k \in \eta_i}\displaystyle\sum_{l \in \eta_i}\left[ \sigma_{ij}\sigma_{kl} + \sigma_{ik}\sigma_{jl} + \sigma_{il}\sigma_{jk} \right] \right. \nonumber \\ \nonumber \\ & \left. + \beta^{2}\displaystyle\sum_{j \in \eta_i}\displaystyle\sum_{k \in \eta_i}\displaystyle\sum_{l \in \eta_i}\displaystyle\sum_{m \in \eta_i}\left[ \sigma_{jk}\sigma_{lm} + \sigma_{jl}\sigma_{km} + \sigma_{jm}\sigma_{kl} \right] \right\}  \label{eq:var} 
\end{align} which in the matrix-vector notation is given by:

\begin{align}   
	\upsilon_{\beta} & = \frac{ \sigma^2 \left\|\Sigma_{p}^{-}\right\|_{+} + 2 \left\|\vec{\rho}\otimes\vec{\rho}^T\right\|_{+} - 6\beta \left\| \vec{\rho}^{T} \otimes \Sigma_{p}^{-} \right\|_{+} + 3\beta^2 \left\| \Sigma_{p}^{-} \otimes \Sigma_{p}^{-} \right\|_{+} }{\left\| \Sigma_{p}^{-} \right\|_{+}^{2}} = \\ \nonumber \\ \nonumber & = \frac{\sigma^{2}}{\left\|\Sigma_{p}^{-}\right\|_{+}} +  \frac{\sigma^{4}\Delta_{\beta} \left( \vec{\rho}, \Sigma_{p}^{-} \right)}{\left\|\Sigma_{p}^{-}\right\|_{+}^{2}} = \frac{1}{\Psi_{\beta}} + \frac{1}{\Psi_{\beta}^{2}}\left( \Phi_{\beta} - \Psi_{\beta} \right) %= \frac{1}{\Psi_{\beta}}\left[ 1 + \frac{1}{\Psi_{\beta}}\left(\Phi_{\beta} - \Psi_{\beta} \right) \right]
\end{align}
\vspace{0.2cm}

Note that when information equilibrium prevails, that is, $\Phi_{\beta} = \Psi_{\beta}$, the asymptotic variance is given by the inverse of the expected Fisher information. However, the interpretation of this equation indicates that the uncertainty in the estimation of the inverse temperature parameter is minimized when $\Psi_{\beta}$ is maximized. Essentially, it means that in average the local pseudo-likelihood functions are not flat, that is, small changes on the local configuration patterns along the system cannot cause abrupt changes in expected global behavior (the global spatial dependence struture is not susceptible to sharp changes). To reach this condition there must be a reasonable degree of agreement between the neighboring elements throughout the system, a behavior that is usually associated to low temperature states ($\beta$ is above a critical value and there is a visible induced spatial dependence struture).

\section{The Fisher Curve}

With the definition of $\Phi_{\beta}$, $\Psi_{\beta}$ and $H_{\beta}$ we have the necessary tools to compute three important information-theoretic measures of a global configuration of the system. Our idea is that we can study the behavior of a complex system by constructing a parametric curve in this information-theoretic space as a function of the inverse temperature parameter $\beta$. Our expectation is that the resulting trajectory provides a geometrical interpretation of how the system moves from a initial configuration A (with a low entropy value for instance) to a desired final configuration B (with a greater value of entropy for instance), since the Fisher information plays an important role in providing a natural metric to the Riemannian manifolds of statistical models \cite{Amari, Kass1989}. We will call the path from global state A to global state B as the \emph{Fisher curve} (from A to B) of the system, denoted by $\vec{F}_{A}^{B}(\beta)$. Instead of using the time as parameter to build the curve $\vec{F}$, we parametrize $\vec{F}$ by the inverse temperature parameter $\beta$.

\begin{mydef}
Let an isotropic pairwise GMRF be defined on a lattice $S=\left\{ s_{1}, s_{2}, \ldots , s_{n} \right\}$ with a neighborhood system $\eta_{i}$ and $\mathbf{X}^{(\beta_{1})},\mathbf{X}^{(\beta_{2})},\ldots,\mathbf{X}^{(\beta_{n})}$ be a sequence of outcomes (global configurations) produced by different values of $\beta_{i}$ (inverse temperature parameters) for which $A = \beta_{MIN} = \beta_{1} < \beta_{2} < \cdots < \beta_{n} = \beta_{MAX} = B$. The system's Fisher curve from $A$ to $B$ is defined as the function $\vec{F}:\Re \rightarrow \Re^{3}$ that maps each configuration $\mathbf{X}^{(\beta_{i})}$ to a point $\left( \Phi_{\beta_{i}}, \Psi_{\beta_{i}}, H_{\beta_{i}} \right)$ from the information space, that is:
\end{mydef}

\begin{equation}
	\vec{F}_{A}^{B}\left(\beta \right) = \left( \Phi_{\beta}, \Psi_{\beta}, H_{\beta} \right) \qquad\qquad \beta = A,\ldots,B
\end{equation} where $\Phi_{\beta}$, $\Psi_{\beta}$ and $H_{\beta}$ denote the type-I expected Fisher information, the type-II expected Fisher information and the Shannon entropy of the global configuration $\mathbf{X}^{(\beta)}$, respectively.

%Assuming that $\mathbf{X^{(\beta)}}=\{x_{1}^{(\beta)}, x_{2}^{(\beta)}, \ldots, x_{n}^{(\beta)} \}, \beta = \beta_{MIN},\ldots,\beta_{MAX}$ denotes the global configuration of the system for each $\beta$, the Fisher curve of the system is defined by the function that maps each  $\vec{F}:\Re \rightarrow \Re^{3}$ for $\beta \in [\beta_{MIN},\beta_{MAX}]$:

  In the next Sections we show some computational experiments that illustrate the effectiveness of the proposed tools in measuring the information encoded in complex systems. We want to investigate what happens to the Fisher curve as the inverse temperature parameter is modified in order to control the system's global behavior. Our main conclusion, which is supported by experimental analysis, is that $\vec{F}_{A}^{B}(\beta) \neq \vec{F}_{B}^{A}(\beta)$. In other words, in terms of information, moving towards higher entropy states is not the same as moving towards lower entropy states, since the \emph{Fisher curves} that represents the trajectory between the initial state A and the final state B are significantly different.

\section{Computational simulations}

   This Section discusses some numerical experiments proposed to illustrate some applications of the derived tools in both simulations and real data. Our computational investigations were divided in two main sets of experiments:
   
\begin{enumerate}
	\item Static data: analysis of the local and global versions of the measures ($\phi_{\beta}$, $\psi_{\beta}$, $\Phi_{\beta}$, $\Psi_{\beta}$ and $H_{\beta}$) in both simulated and real data considering a fixed inverse temperature parameter;
	\item Dynamic data: analysis of the global versions of the measures ($\Phi_{\beta}$, $\Psi_{\beta}$ and $H_{\beta}$) along Markov Chain Monte Carlo (MCMC) simulations in which the inverse temperature parameter is modified to control the expected global behavior;
	%\item analysis of the global versions of the measures $\Phi_{\beta}$, $\Psi_{\beta}$ and $H_{\beta}$ in the context of image processing;
\end{enumerate}

\subsection{Learning from Static Data with Information-Theoretic Measures}   

\begin{figure}[!ht]
\begin{center}
\includegraphics[scale=0.65]{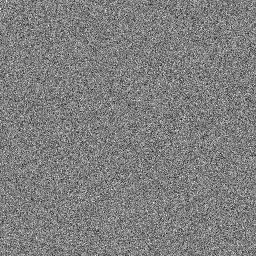}
\includegraphics[scale=0.86]{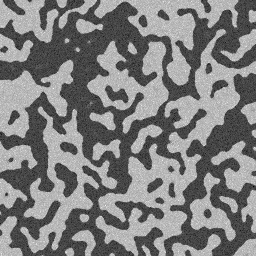}
\end{center}
\caption{
{\bf Example of GMRF model outputs.} The values of the inverse parameter $\beta$ in the left and right images are 0 and 0.125, respectively.
}
\label{fig:res1}
\end{figure}

	First, in order to illustrate the application of both forms of local observed Fisher information, $\phi_{\beta}$ and $\psi_{\beta}$, we performed a simple experiment using some synthetic images generated by the Metropolis-Hastings algorithm. The initial configuration is random and after a fixed number of steps, the algorithm produces a valid outcome of an isotropic pairwise GMRF model. Figure \ref{fig:res1} shows an example of initial condition and the resulting outcome considering a second order neighborhood system (8 nearest neighbors). The parameters settings were: $\mu = 0$, $\sigma^{2} = 5$ and $\beta = 0.125$. The number of iterations considered in this MCMC simulation was 1000.

\begin{figure}[!ht]
\begin{center}
\includegraphics[scale=0.65]{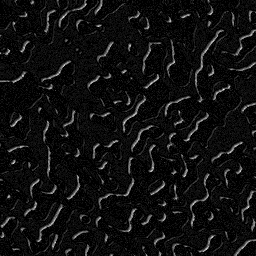}
\includegraphics[scale=0.65]{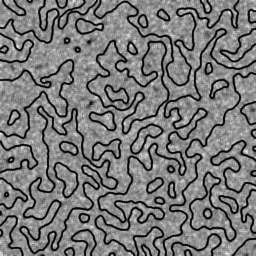}
\includegraphics[scale=0.65]{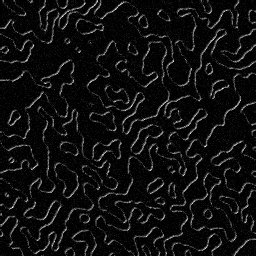}
\end{center}
\caption{
{\bf Fisher information maps.} The first and second information maps were generated by computing $\phi_{\beta}$ and $\psi_{\beta}$ for each observation in the lattice. The third map was produced by computing the local \emph{L-information}, that is, the ratio between the local information measures.
}
\label{fig:res2}
\end{figure}

Three Fisher information maps were generated from the resulting synthetic image. The first one was obtained by calculating the value of type-I observed local Fisher information, $\phi_{\beta}$, for every observation of the system. Similarly, the second one was obtained by using the type-II observed local Fisher information, $\psi_{\beta}$. For the last information map, we used the ratio between $\phi_{\beta}$ and $\psi_{\beta}$, motivated by the fact that boundaries are often composed by patterns that are not expected to be ``aligned'' to the global behavior (high values of $\phi_{\beta}$) and also are somehow unstable (low values of $\psi_{\beta}$). We named this measure, $\phi_{\beta}/\psi_{\beta}$, \emph{L-information}, since it is defined in terms of the first two derivatives of the logarithm of local likelihood function. Figure \ref{fig:res2} shows the obtained information maps as images. Note that while $\phi_{\beta}$ has a strong response for boundaries (the edges are light), $\psi_{\beta}$ has a weak one (edges are dark), an evidence in favor of considering \emph{L-information} in boundariy detection procedures.

\begin{figure}[!ht]
\begin{center}
\includegraphics[scale=0.5]{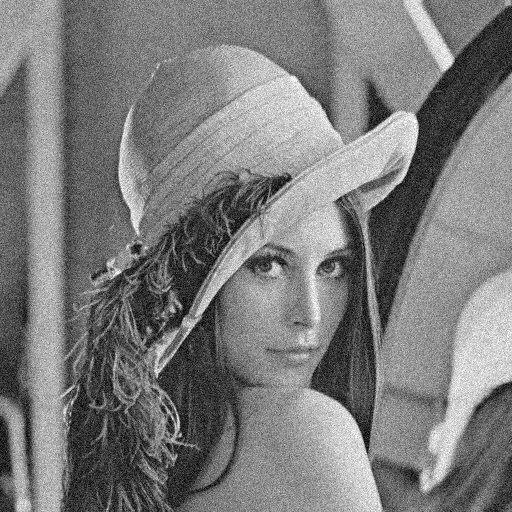}
\includegraphics[scale=0.5]{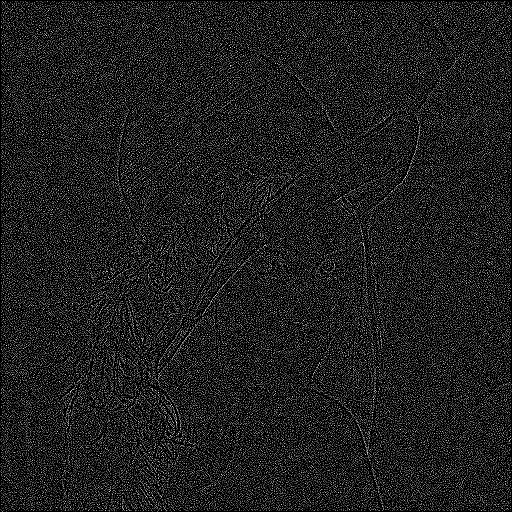} \vspace{0.25cm}
\includegraphics[scale=0.5]{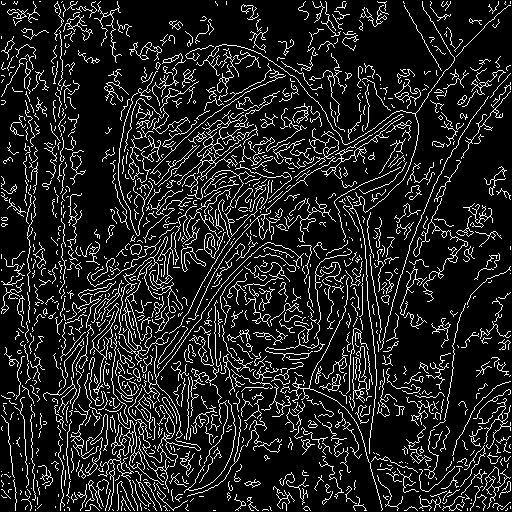}
\includegraphics[scale=0.5]{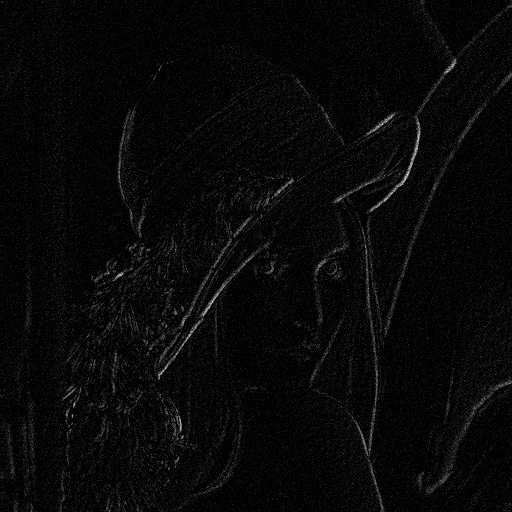}
\end{center}
\caption{
{\bf Edge detection performed in a noisy image.} The results from top to bottom and left to right show the input noisy configuration, the result of the Laplacian filter, the result of the Canny filter and the L-information map ($\hat{\beta}_{MPL} = 0.1140$), respectively. 
}
\label{fig:res3}
\end{figure}

The same experiment was repeated for real image data. Grayscale images were corrupted by additive gaussian to make the edge detection process a harder task. It is known from image processing literature that the problem of detecting egdes in the presence of noise data is extremelly challenging, since typical boundary detectors are based on differential operators which causes noise amplification. In this context, we believe that the proposed tools provide a reasonable solution to such cases. Figure \ref{fig:res3} shows a noisy input image, the solution of the Laplacian edge detector (a usual filter used to detect boundaries in images), the solution of the Canny edge detector (another reference method for boundary detection in images) \cite{Canny} and the respective \emph{L-information} map. Note that the response of the \emph{L-information} map gives a good approximation to the image boundaries, even in the presence of random noise and perturbations. Note also that the \emph{L-information} map retains relevant image information without an excessive smoothing (loss of fine details), which is a positive characteristic for a edge detector filter.

\begin{figure}[!ht]
\begin{center}
\includegraphics[scale=0.5]{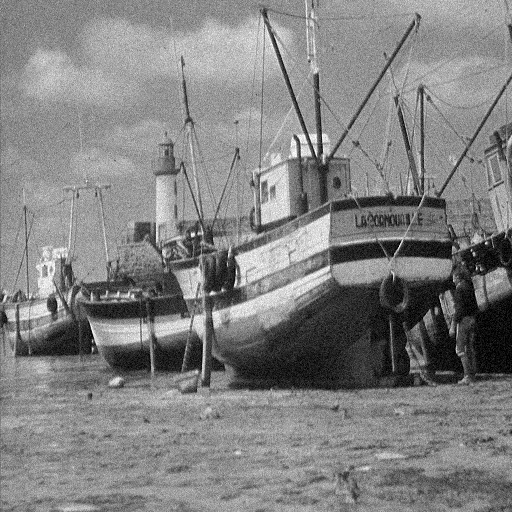}
\includegraphics[scale=0.5]{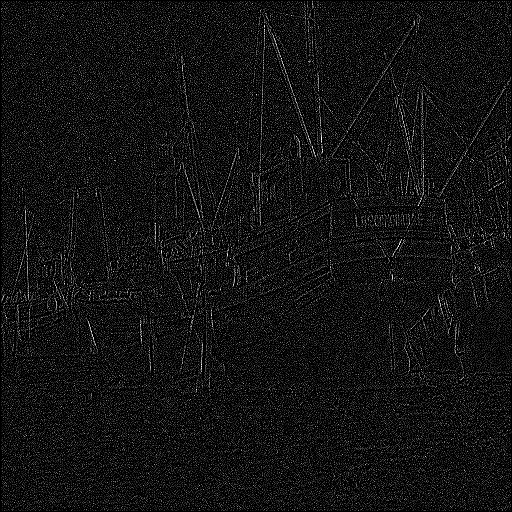} 
\includegraphics[scale=0.5]{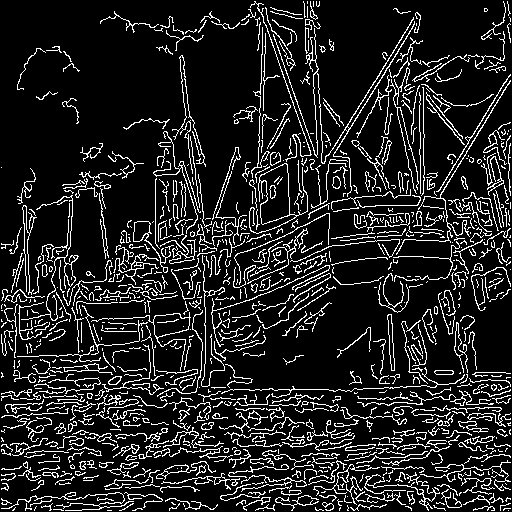}
\includegraphics[scale=0.5]{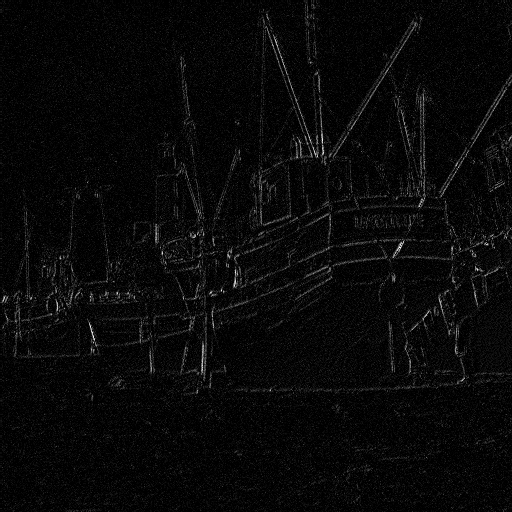}
\end{center}
\caption{
{\bf Edge detection performed in a noisy image.} The results from top to bottom and left to right show the input noisy configuration, the result of the Laplacian filter, the result of the Canny filter and the L-information map ($\hat{\beta}_{MPL} = 0.1266$), respectively. 
}
\label{fig:res4}
\end{figure}

Another similar result considering a different noisyimage can be seen in Figure \ref{fig:res4}. Basically, the same methodology described in the previous experiment was adopted here. Again, the proposed tools performed well and a reasonable amount of relevant information could be extracted by measuring the local Fisher information.

\begin{figure}[!ht]
\begin{center}
\includegraphics[scale=0.4]{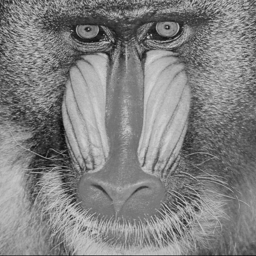}
\includegraphics[scale=0.4]{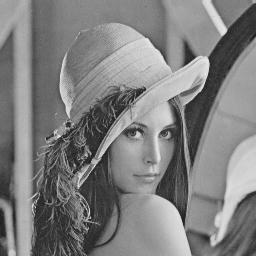} 
\includegraphics[scale=0.4]{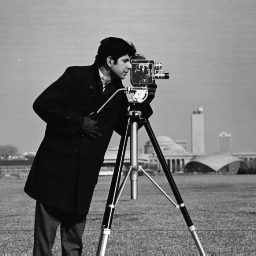}
\includegraphics[scale=0.16]{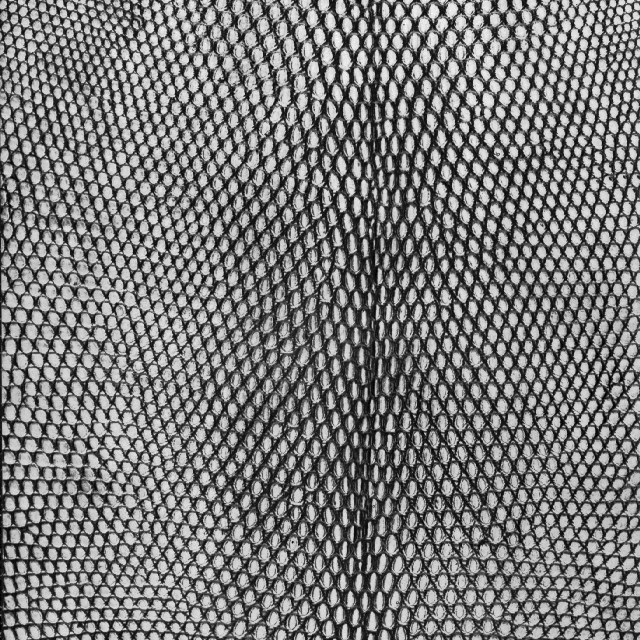}
\end{center}
\caption{
{\bf Measures of information in grayscale images.} From left to right and top to bottom, the of $H_{\beta}$ are: 4.6363, 4.7857, 5.0825 and 5.1590, respectively. Similarly, for $\Phi_{\beta}$, the values are: 8.112, 2.8516, 3.077 and 5.6932. Finally, for $\Psi_{\beta}$ the values are: 51.101, 58.0462, 58.3909 and 19.2041. The image entropy computed by estimating the probabilities from the data histogram shows the values: 7.2279, 7.4227, 7.0097 and 6.2418, respectively. Note that $\Psi_{\beta}$ in the texture piece is an outlier.
}
\label{fig:resbloco}
\end{figure}

To measure the entropy in isotropic pairwise GMRF models we show an illustrative example using some real data in the form of grayscale images. For this experiment four different classical images were considered - Baboon, Lena, Cameraman and a texture piece. Our objective is to investigate how the entropy in GMRF model could be used to quantify and and measure the variability of the local configuration patterns presented in data. Figure \ref{fig:resbloco} shows the values of Fisher information and Shannon entropy for each one of the four images. The results indicate that the Baboon image has the lowest entropy and the texture piece has the highest value. We observed that it is the opposite of what happens if we discard the dependence struture between the observations by setting $\beta = 0$ (that is, for independent observations). The usual image entropy $H$, computed directly from the image histogram provides a completely different information since it relies only in the individual pixel intensities. With the definition of $H_{\beta}$ for the GMRF model it is possible to analyze entropy in a different scale level. Note also that im terms of $\Psi_{\beta}$, the texture image can be considered as an outlier (it shows a significantly smaller value in comparison to other three natural images).

\begin{figure}[!ht]
\begin{center}
\includegraphics[scale=0.7]{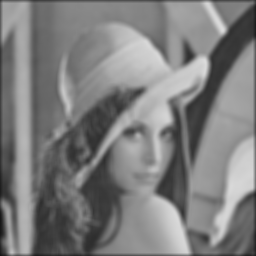}
\includegraphics[scale=0.7]{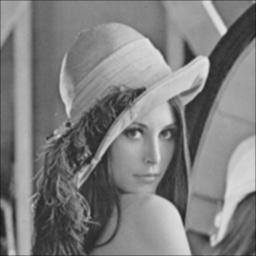} 
\includegraphics[scale=0.7]{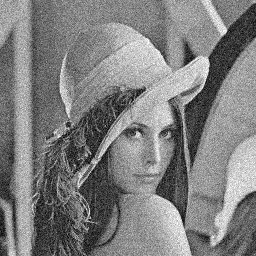}
\includegraphics[scale=0.7]{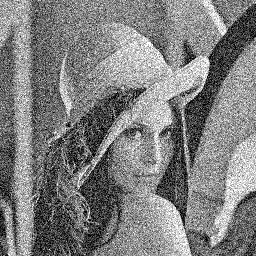}
\end{center}
\caption{
{\bf Information-theoretic measures for different versions of the Lena image.} From left to right and top to bottom, the results for $H_{\beta}$ are: 4.6579, 4.7243, 4.9001 and 5.2040. Similarly, the results for $\Phi_{\beta}$ are: 0.0580, 0.4776, 8.7643 and 17.7712. Finally, the results for $\Psi_{\beta}$ are: 62.7282, 60.9517, 52.6158, 39.3270. Note that $\Phi_{\beta}$ is a good measure in capturing the differences between the images.
}
\label{fig:resruido}
\end{figure}

Finally, to investigate how these information-theoretic measures are related to the distribution of patterns along an isotropic pairwise GMRF model we compared  the values of $\Phi_{\beta}$, $\Psi_{\beta}$ and $H_{\beta}$ for different versions of the same grayscale image, from a very blurred one (less variability of local patterns) to a very noisy one (more variability of local patterns). Figure \ref{fig:resruido} shows the obtained results. Note that the uncertainty about the real value of the inverse temperature parameter $\beta$ grows as the noise level increases since $\Psi_{\beta}$ is significantly reduced. Note also that $\Phi_{\beta}$ is an effective measure in capturing the differences between the images as they get smoother or noisier.

\vspace{1cm}

\subsection{Learning from Dynamic Systems with Information-Theoretic Measures}

In order to study the behavior of a complex system that evolves from an initial state A to another state B, we used the Metropolis-Hastings algorithm, a MCMC simulation method, to generate a sequence of valid isotropic pairwise GMRF model outcomes for different values of the inverse temperature parameter $\beta$. The purpose of the experiment is to observe what happens to $\Phi_{\beta}$, $\Psi_{\beta}$ and $H_{\beta}$ when the system evolves from a random initial state to other configurations. In other words, we want to investigate the \emph{Fisher curve} of the system in order to characterize its behavior in the information space. Basically, the idea is to use the \emph{Fisher curve} as a signature for the expected behavior of a system modeled by an isotropic pairwise GMRF. 
	
\begin{figure}[!ht]
\begin{center}
\includegraphics[scale=0.4]{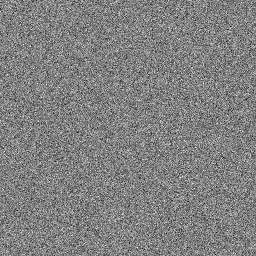}
\includegraphics[scale=0.4]{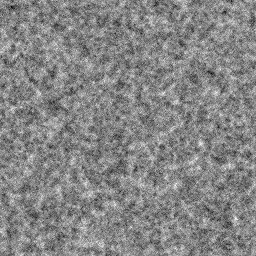}
\includegraphics[scale=0.4]{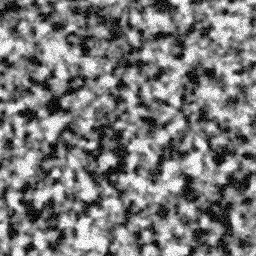}
\includegraphics[scale=0.4]{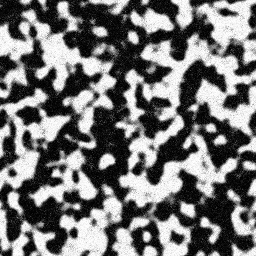}
\includegraphics[scale=0.4]{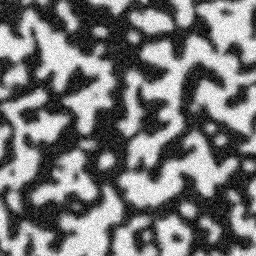}
\includegraphics[scale=0.4]{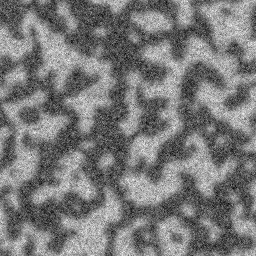}
\includegraphics[scale=0.4]{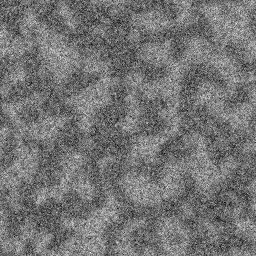}
\includegraphics[scale=0.4]{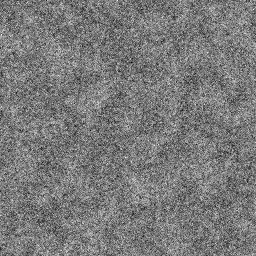}
\end{center}
\caption{
{\bf Global configurations along a MCMC simulation.} Evolution of the global state as the inverse temperature parameter $\beta$ is modified to control the system's behavior.
}
\label{fig:res5}
\end{figure}

	To simulate a system where we can control the inverse temperature parameter, we define an updating rule for $\beta$ based on fixed increments. In summary, we start with a minimum value $\beta_{MIN}$. Then, the value of $\beta$ in the iteration $t$ is defined as the value of $\beta$ in $t-1$ plus a small increment ($\Delta\beta$), until it reaches a pre-defined upper bound $\beta_{MAX}$. The process in then repeated with negative increments $-\Delta\beta$, until the inverse temperature reaches its minimum value $\beta_{MIN}$ again. This process continues for a fixed number of iterations $N_{MAX}$ during a MCMC simulation. As a result of this approach, a sequence of GMRF samples is produced. We use this sequence to calculate $\Phi_{\beta}$, $\Psi_{\beta}$ and $H_{\beta}$ and define the \emph{Fisher curve} $\vec{F}$ for $\beta =\beta_{MIN},\ldots,\beta_{MAX}$. Figure \ref{fig:res5} shows some of the system's configurations along a MCMC simulation. In this experiment, the parameters were defined as: $\beta_{MIN} = 0$, $\Delta\beta = 0.001$, $\beta_{MAX} = 0.15$ and $N_{MAX} = 1000$, $\mu = 0$, $\sigma^{2} = 5$ and $\eta_{i} = \{(i-1,j-1), (i-1,j), (i-1,j+1), (i,j-1), (i,j+1), (i+1,j-1), (i+1,j), (i+1,j+1)\}$.

\begin{figure}[!ht]
\begin{center}
\includegraphics[scale=0.7]{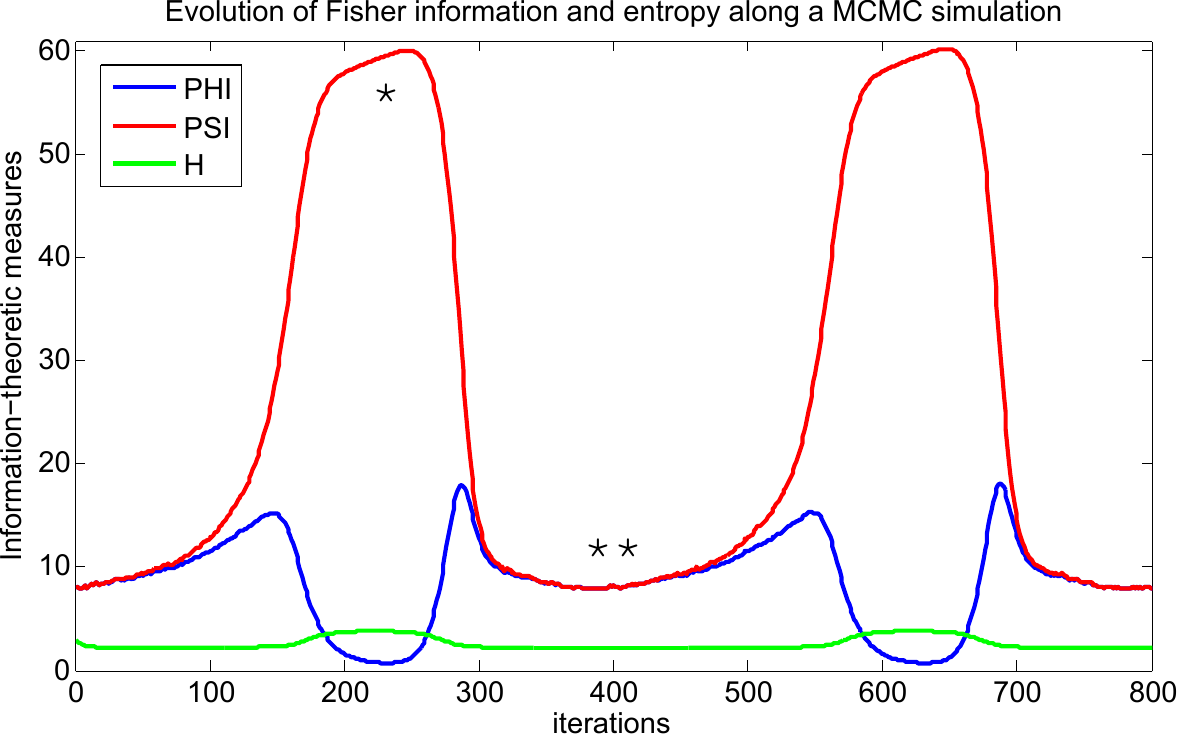}  % Com Inkscape ficou perfeito!
\end{center}
\caption{
{\bf Evolution of Fisher information along a MCMC simulation.} As the difference between $\Phi_{\beta}$ and $\Psi_{\beta}$ is maximized (*), the uncertainty about the real inverse temperature parameter is minimized and the number of informative patterns increases. In the information equilibrium condition (**) it is hard to find informative patters since there is no induced spatial dependence structure.
}
\label{fig:res6}
\end{figure}

A plot of both forms of the expected Fisher information, $\Phi_{\beta}$ and $\Psi_{\beta}$, for each iteration of the MCMC simulation is shown in Figure \ref{fig:res6}. The graph produced by this experiment show some interesting results. First of all, regarding upper and lower bounds on these measures. It is possible to note that when there is no induced spatial dependence structure ($\beta \approx 0$), we have an information equilibrium condition ($\Phi_{\beta} = \Psi_{\beta}$ and the information equality holds). In this condition the observations are practically independent in the sense that all local configuration patterns convey approximately the same amount of information. Thus, it is hard to find and separate the two categories of patterns we know: the informative and the non-informative ones. Once they all behave in a similar manner, there is no informative pattern to highlight. Moreover, in this information equilibrium situation, $\Psi_{\beta}$ reaches its lower bound (in this simulation we observed that in the equilibrium $\Phi_{\beta} \approx \Psi_{\beta} \approx 8$), indicating that this condition emerges when the system is most susceptible to a change in the expected global behavior, since the uncertainty about $\beta$ is maximum at this moment. In other words, modification in the behavior of a small subset of local patterns may guide the system to a totally different stable configuration in the future. 

The results also show that the difference between $\Phi_{\beta}$ and $\Psi_{\beta}$ is maximum when the system operates with large values of $\beta$, that is, when organization emerges and there is a strong dependence struture among the random variables (the global configuration shows clear visible clusters and boundaries between them). In such states, it is expected that the majority of patterns be aligned to the global behavior, which causes the appearance of few but highly informative patterns: those connecting elements from different regions (boundaries). Besides that, the simulation suggests that it takes more time for the system to go from the information equilibruim state to organization than the opposite. We will see how this fact becomes clear by analyzing the \emph{Fisher curve} of the system.  Finally, the results also suggest that both $\Phi_{\beta}$ and $\Psi_{\beta}$ are bounded by a superior value, possibly related to the size of the neighborhood system.

% PDF realmente ficou com qualidade superior. Porém precisa recortar a área em branco ao redor da imagem

\begin{figure}[!ht]
\begin{center}
\includegraphics[scale=0.6]{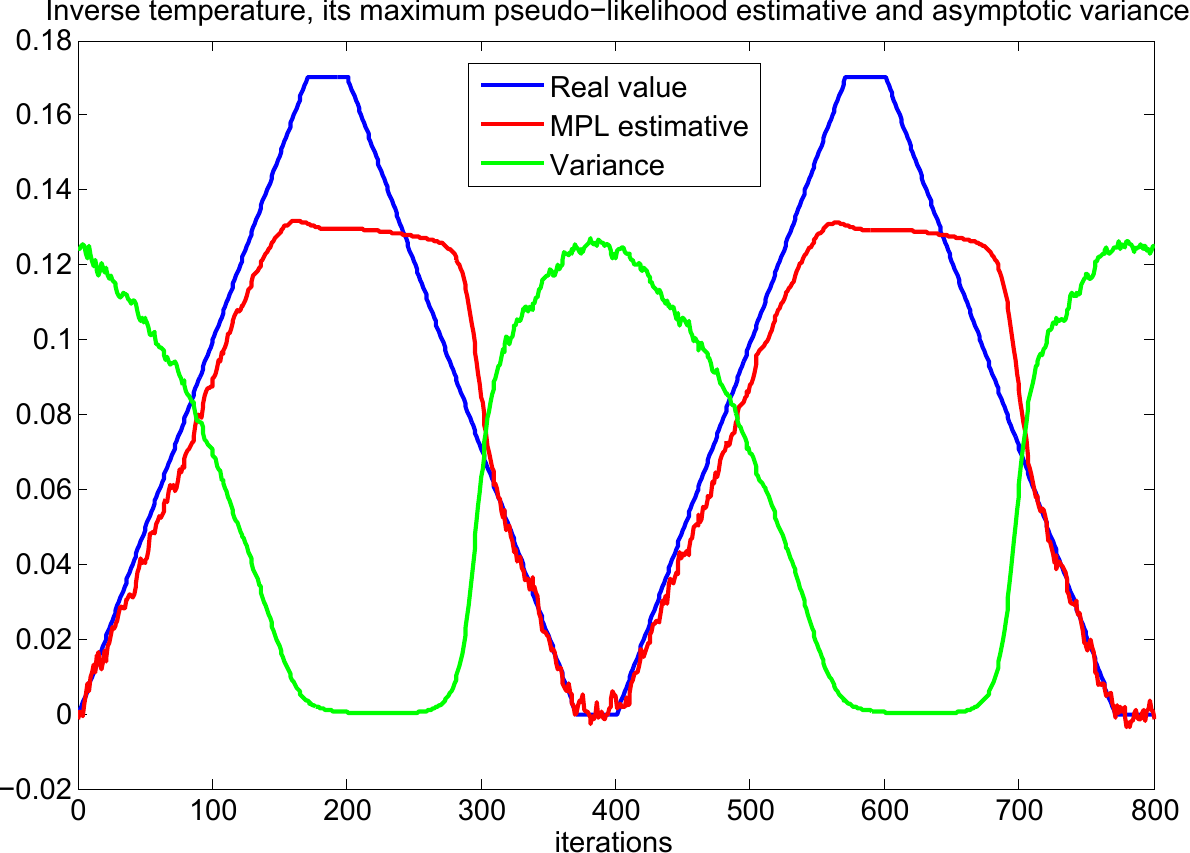}  
\end{center}
\caption{
{\bf Real and estimated inverse temperatures along the MCMC simulation.} The system's global behavior is controled by the real inverse temperature parameter values (blue line), used to generate the GMRF outputs. The maximum pseudo-likelihood estimative is used to compute both $\Phi_{\beta}$ and $\Psi_{\beta}$. Note that the uncertainty about the inverse temperature increases as $\beta \rightarrow 0$ and the system approaches the information equilibrium condition.
}
\label{fig:variancia}
\end{figure}

Figure \ref{fig:variancia} shows the real parameter values used to generate the GMRF outputs (blue line), the maximum pseudo-likelihood estimative used to calculate $\Phi_{\beta}$ and $\Psi_{\beta}$ (red line), and also a plot of the asymptotic variances (uncertainty about the inverse temperature) along the entire MCMC simulation.

\begin{figure}[!ht]
\begin{center}
\includegraphics[scale=0.6]{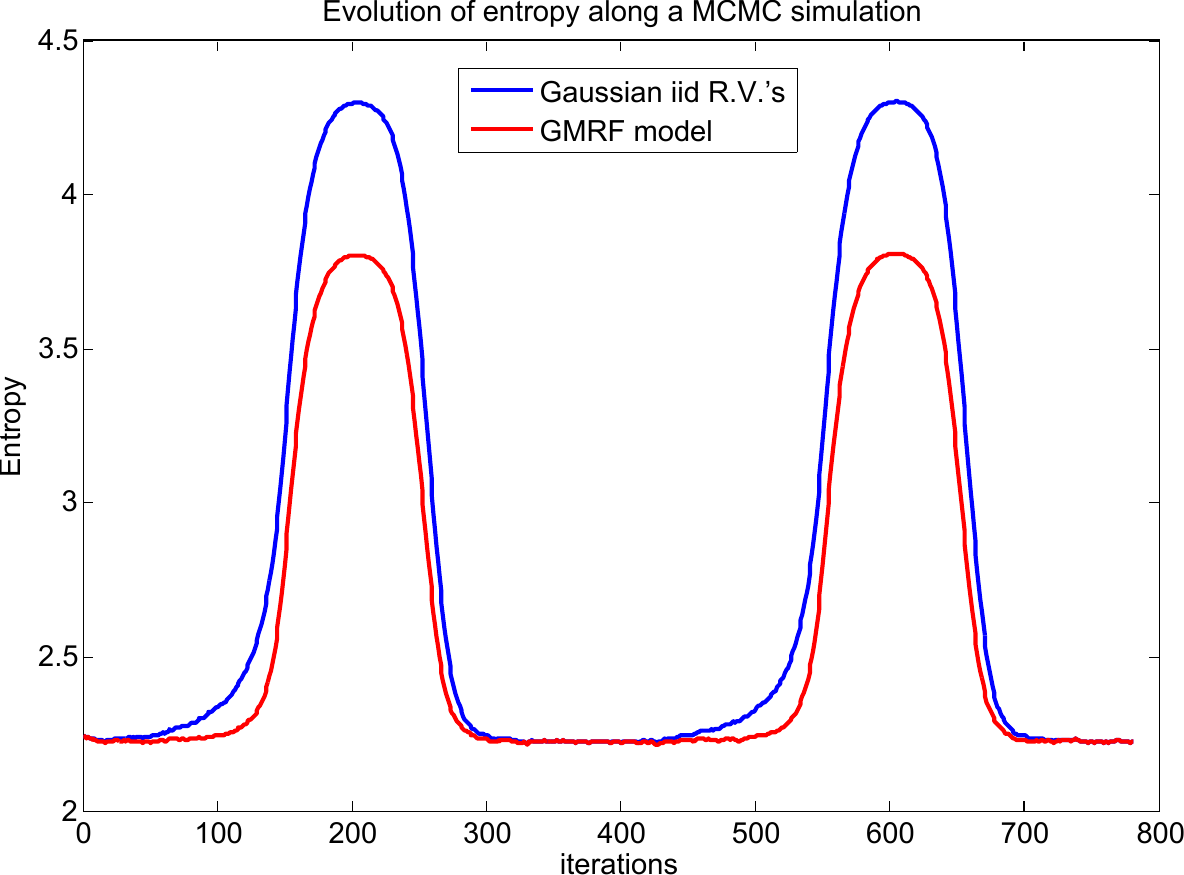}
\end{center}
\caption{
{\bf Evolution of Shannon entropy along a MCMC simulation.} $H_{\beta}$ start to grow when the system leaves the equilibrium condition, where the entropy in the isotropic pairwise GMRF model is identical to the entropy of a simple Gaussian random variable (since $\beta \rightarrow 0$).
}
\label{fig:res7}
\end{figure}

We now proceed to the analysis of the Shannon entropy of the system along the simulation. Despite showing a behavior similar to $\Psi_{\beta}$, the range of values for entropy is significantly smaller. In this simulation we observed that $0 \leq H_{\beta} \leq 4.5$, $0 \leq \Phi_{\beta} \leq 18$ and $8 \leq \Psi_{\beta} \leq 61$. An interesting point is that knowledge of $\Phi_{\beta}$ and $\Psi_{\beta}$ allows us to infer the entropy of the system. For example, looking at Figures \ref{fig:res6} and \ref{fig:res7} we can see that $\Phi_{\beta}$ and $\Psi_{\beta}$ start to diverge a little bit earlier ($t \approx 80$) than the entropy in a GMRF model begins to grow ($t \approx 120$). Therefore, in an isotropic pairwise GMRF model, if the system is close to the information equilibrium condition, then $H_{\beta}$ is low since there is little variability in the observed configuration patterns. When the difference between $\Phi_{\beta}$ and $\Psi_{\beta}$ is large, $H_{\beta}$ increases.  %Note also that $H_{\beta}$ shows a more symmetric behavior in comparison to $\Phi_{\beta}$ or $\Psi_{\beta}$.

\begin{figure}[!ht]
\begin{center}
\includegraphics[scale=0.6]{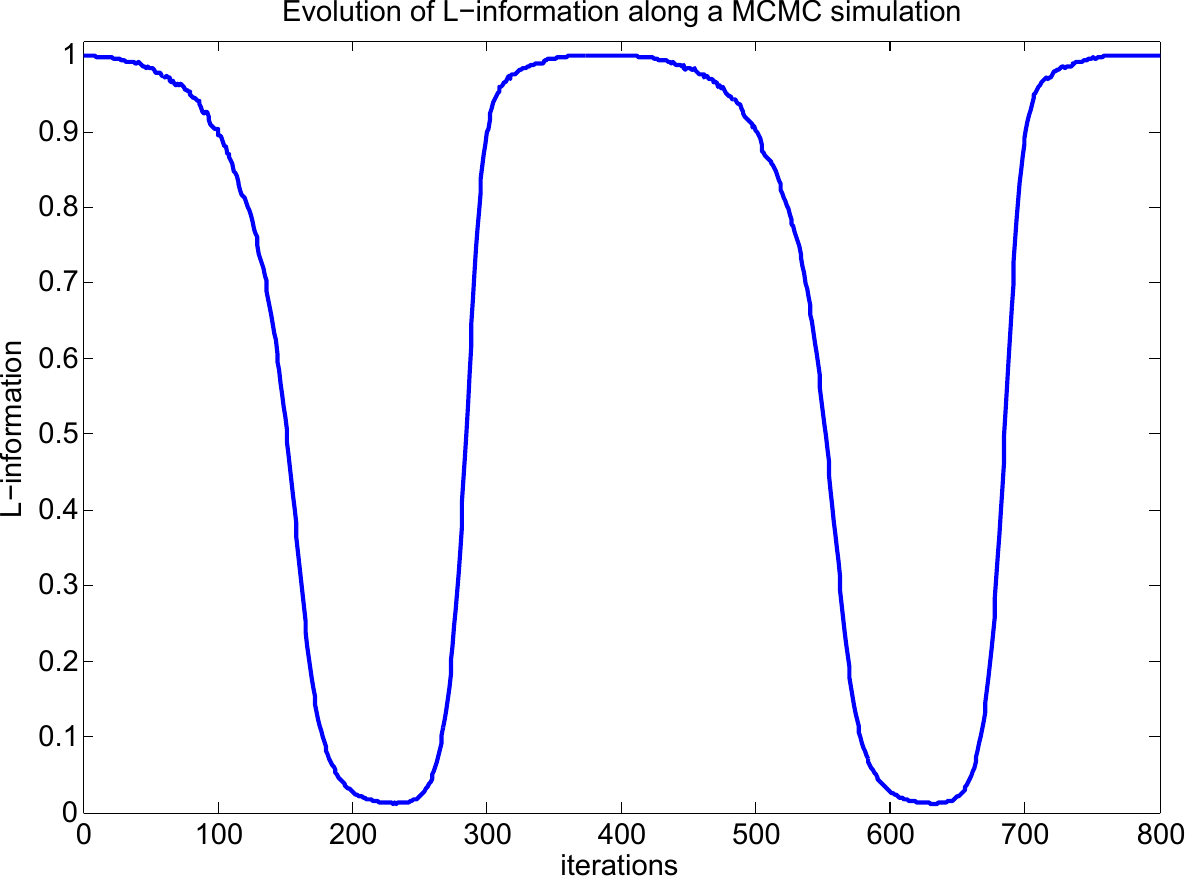}
\end{center}
\caption{
{\bf Evolution of \emph{L-information} along a MCMC simulation.}  When $L_{\beta}$ 
approaches 1 the system tends to the information equilibrium condition. 
For values close to 0, the system tends to the maximum entropy condition.
}
\label{fig:res8}
\end{figure}

Another interesting global information-theoretic measure is \emph{L-information}, from now on denote by $L_{\beta}$, since it conveys all the information about the likelihood function (in a GMRF model only the first two derivatives of $L( \vec{\theta}; \mathbf{X}^{(t)} )$ are not null). $L_{\beta}$ is defined as the ratio between the two forms of expected Fisher information, $\Phi_{\beta}$ and $\Psi_{\beta}$. A nice property about this measure is that $0 \leq L_{\beta} \leq 1$. With this single measurement it is possible to gain insights about the global system behavior. Figure \ref{fig:res8} shows that a value close to one indicates a system approximating the information equilibrium condition, while a value close to zero indicates a system close to the maximum entropy condition (a stable configuration with boundaries and informative patterns).

%\begin{figure}
%	\centering
%		\includegraphics[scale=0.6]{./figuras/MCMC_L-information_800it.png}
%	\caption{Evolution of \emph{L-information} along a MCMC simulation where the parameter $\beta$ is modified in order to control the system's behavior. When \emph{L-information} approaches one the system tends to information equilibrium. For values close to zero, different kinds of configuration pattern are observable.}
	%\label{fig:res8}
%\end{figure}
 
\begin{figure}[!ht]
\begin{center}
\includegraphics[scale=0.6]{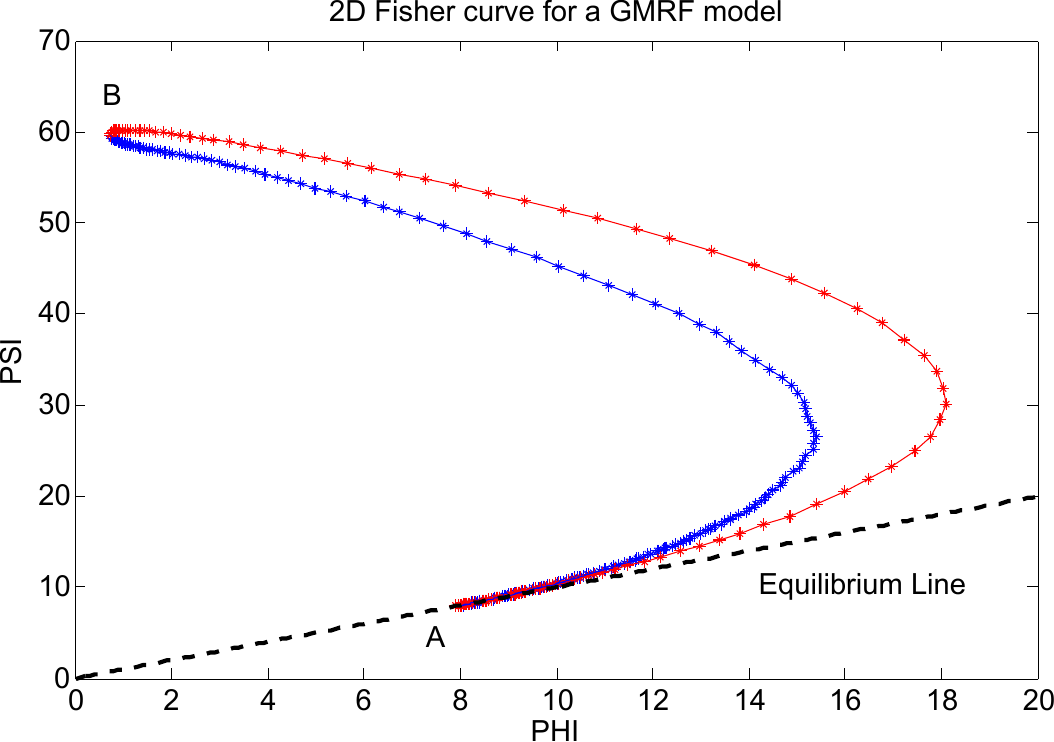}\vspace{1cm}
\includegraphics[scale=0.6]{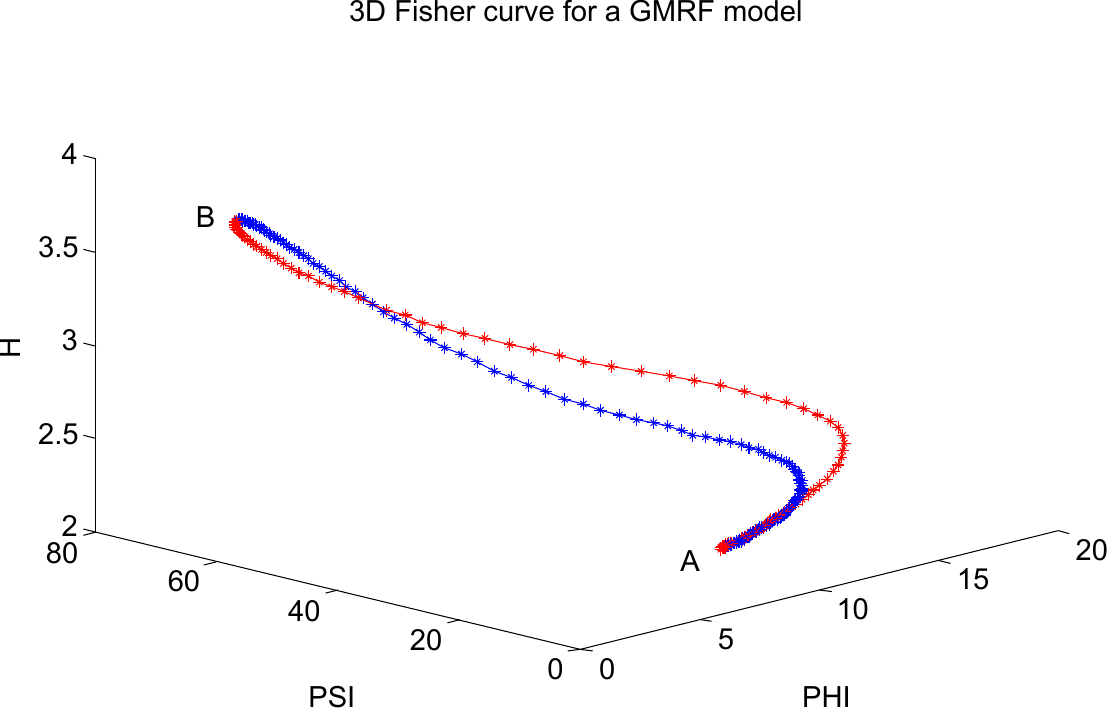}
\end{center}
\caption{
{\bf 2D and 3D Fisher curves of a complex system along a MCMC simulation.} The graph shows a parametric curve obtained by varying the $\beta$ parameter from $\beta_{MIN}$ to $\beta_{MAX}$ and back. Note that, from a differential geometry perspective, as the divergence between $\Phi_{\beta}$ and $\Psi_{\beta}$ increases, the torsion of the parametric curve becomes evident (the curve leaves the plane of constant entropy). 
}
\label{fig:res9}
\end{figure}

\begin{figure}[!ht]
\begin{center}
\includegraphics[scale=0.75]{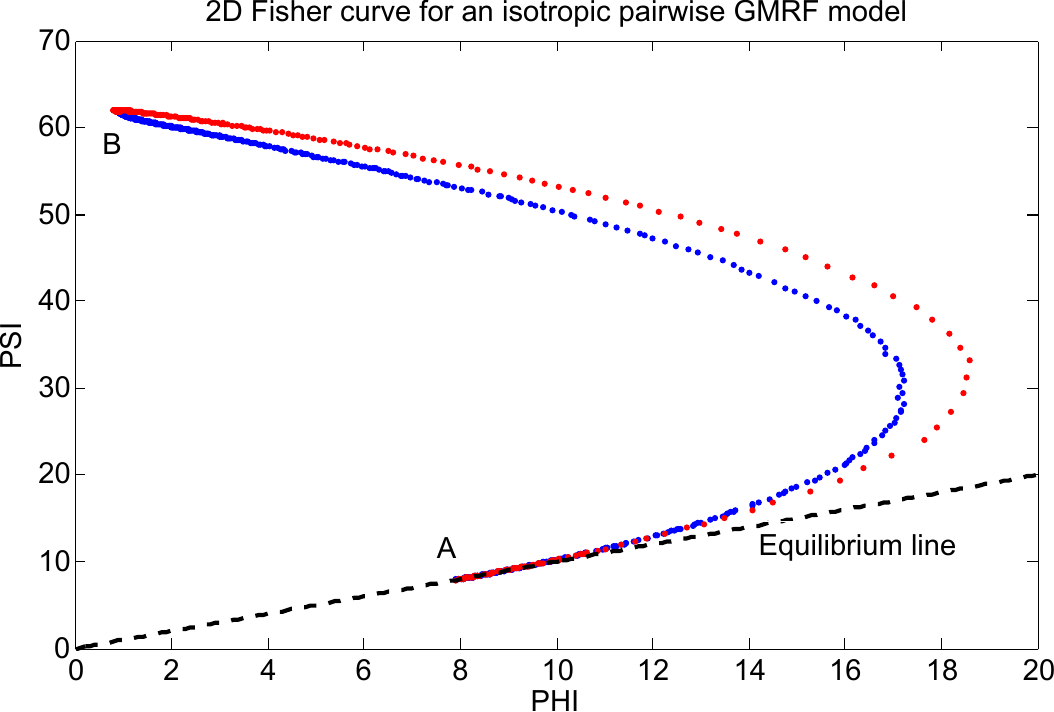}\vspace{1cm}
\includegraphics[scale=0.80]{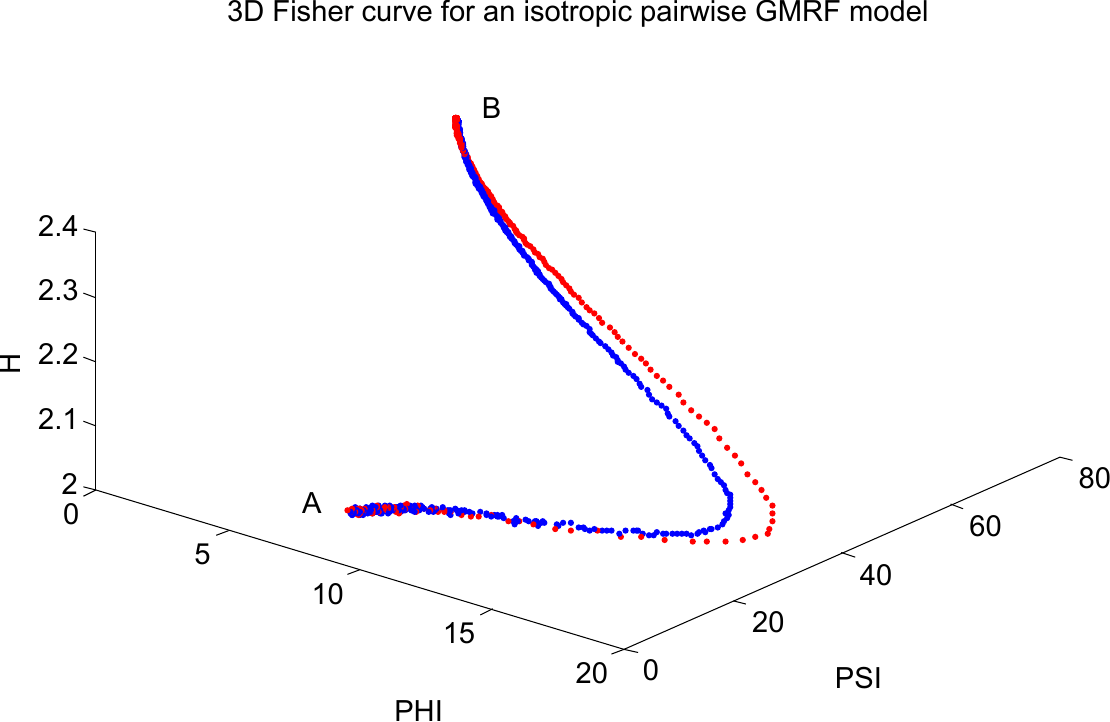}
\end{center}
\caption{
{\bf 2D and 3D Fisher curves along another MCMC simulation.} The graph shows a parametric curve obtained by varying the $\beta$ parameter from $\beta_{MIN}$ to $\beta_{MAX}$ and back. Note that, from a geometrical perspective, the properties of these curves are essentially the same as the ones from the previous simulation. 
}
\label{fig:long}
\end{figure}

%-----> Fisher curves
To investigate the intrinsic non-linear connection between $\Phi_{\beta}$, $\Psi_{\beta}$ and $H_{\beta}$ in a complex system modeled by an isotropic pairwise GMRF model, we now analyze its \emph{Fisher curves}. The first curve, which is a planar one, is defined as $\vec{F}(\beta) = (\Phi_{\beta}, \Psi_{\beta})$, for $A=\beta_{min}$ to $B=\beta_{max}$ and shows how Fisher information changes when the inverse temperature of the system is modified to control the global behavior. Figure \ref{fig:res9} shows the results. In the first image, the blue piece of the curve is the path from A to B, that is, $\vec{F}(\beta)_{A}^{B}$, and the red piece is the inverse path (from B to A), that is, $\vec{F}(\beta)_{B}^{A}$. We must emphasize that $\vec{F}(\beta)_{A}^{B}$ is the trajectory from a lower entropy global configration to a higher entropy global configuration. On the other hand, when the system moves from B to A, we are moving towards entropy minimization. To make this clear, the second image of Figure \ref{fig:res9} illustrates the same \emph{Fisher curve} as before, but now in three dimensions, that is, $\vec{F}(\beta)=( \Phi_{\beta}, \Psi_{\beta}, H_{\beta} )$. For comparison purposes Figure \ref{fig:long} shows the \emph{Fisher curves} for another MCMC simulation with different parameter settings. Note that the shape of the curves are quite similar to those in Figure \ref{fig:res9}.

We can see that the majority of points along the \emph{Fisher curve} is concentrated around two regions of high curvature: A) around the information equilibrium condition (absence of short-term and long-term correlations since $\beta = 0$) and B) around the maximum entropy value, where the divergence between the information values are maximum (self-organization emerges since $\beta$ is greater than a critical value $\beta_{c}$). The points thst lie in the middle of the path connecting these two regions represent the system undergoing a phase transition. Its properties change rapidly and in an assimetric way since $\vec{F}(\beta)_{A}^{B} \neq \vec{F}(\beta)_{B}^{A}$ for a given natural orientation.

\begin{figure}[!ht]
\begin{center}
\includegraphics[scale=0.6]{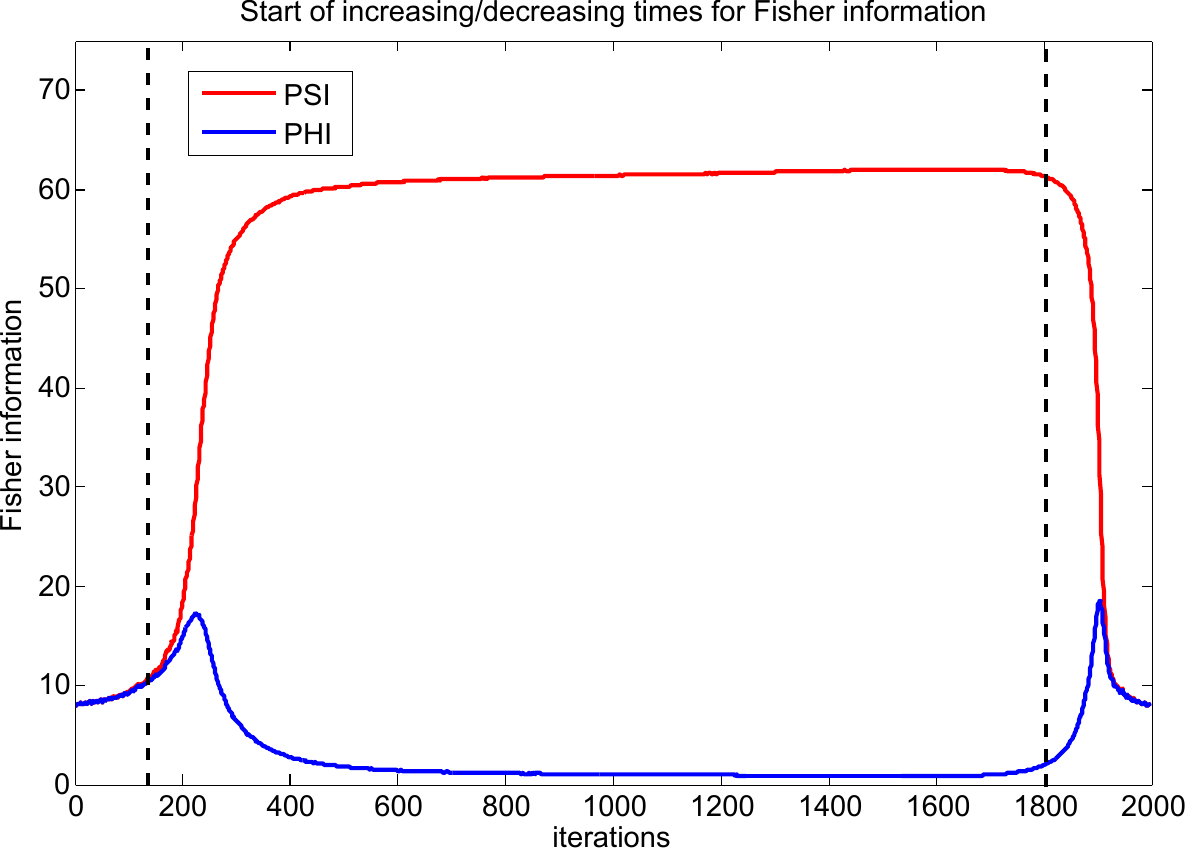}\vspace{1cm}
\includegraphics[scale=0.6]{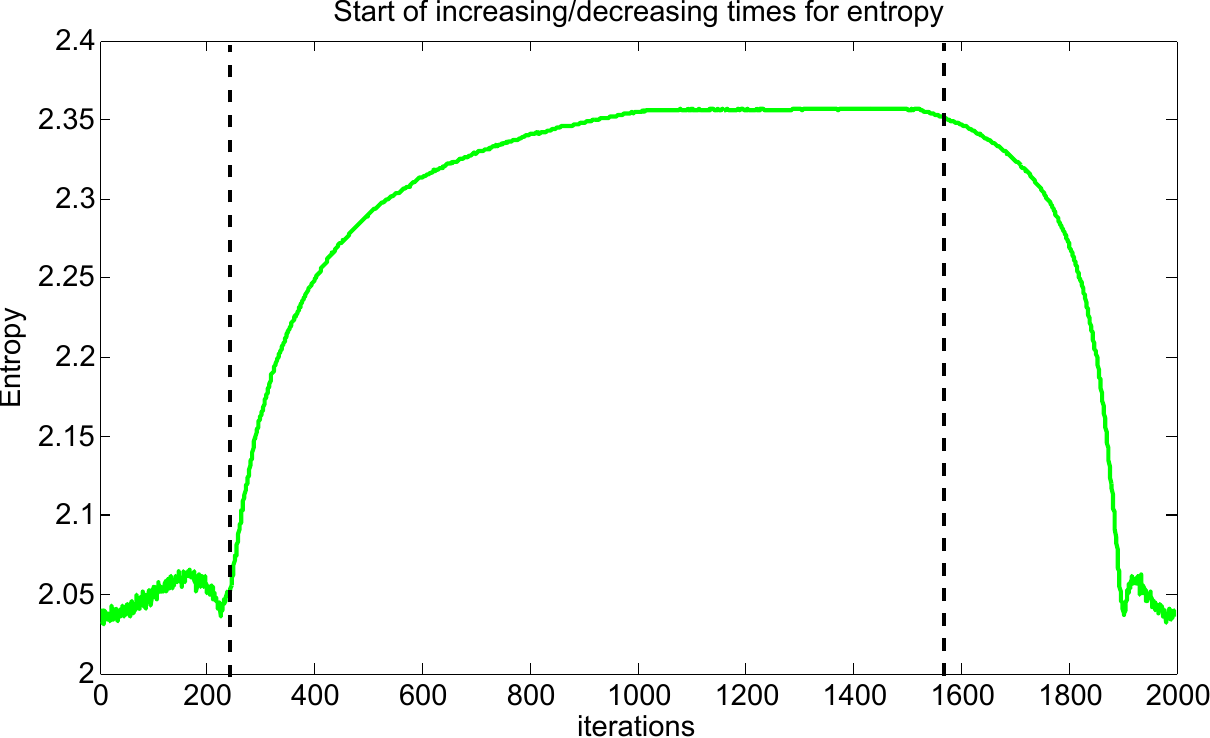}
\end{center}
\caption{
{\bf Relations between entropy and Fisher information.} When a system modeled by an isotropic pairwise GMRF evolves in the natural orientation (forward in time), two rules that relate Fisher information and entropy can be observed: 1. Increase in $\Psi_{\beta}$ is the cause to an increase in $H_{\beta}$ (the increase in $H_{\beta}$ is a consequence of the increase in $\Psi_{\beta}$); 2. Decrease in $H_{\beta}$ is the cause to a decrease in $\Psi_{\beta}$ (the decrease in $\Psi_{\beta}$ is a consequence of the decrease in $H_{\beta}$). In other words, when moving towards higher entropy states, changes in Fisher information preceeds changes in entropy ($\Psi_{\beta}$ ``pushes $H_{\beta}$ up''). When moving towards lower entropy states changes in entropy preceeds changes in Fisher information ($H_{\beta}$ ``pushes $\Psi_{\beta}$ down'').}
\label{fig:increase}
\end{figure}

By now, some observations can be highlighted. First, the natural orientation of the \emph{Fisher curve} defines the direction of time. The natural A-B path (increase in entropy) is given by the blue curve and the natural B-A path (decrease in entropy) is given by the red curve. In other words, the only possible way to walk from A to B (increase $H_{\beta}$) by the red path or to walk from B to A (decrease $H_{\beta}$) by the blue path would be moving back in time (by running the recorded simulation backwards). Eventually, we believe that a possible explanation for this fact could be that those blue and red paths defined by the \emph{Fisher curves} $\vec{F}(\beta)_{A}^{B}$ and $\vec{F}(\beta)_{B}^{A}$ are part of a non orientable manifold, such as a Möbius strip in which the edge is irregular. Thus, even the basic notion of time seems to be deeply connected with the relationship between entropy and Fisher information in a complex system: in the natural orientation (forward in time), it seems that the divergence between $\Phi_{\beta}$ and $\Psi_{\beta}$ is the cause of an increase in the entropy, and the decrease of entropy is the cause of the convergence of $\Phi_{\beta}$ and $\Psi_{\beta}$. During the experimental analysis, we repeated the MCMC simulations with different parameters settings and the observed behavior for Fisher information and entropy was the same. Figure \ref{fig:increase} shows the graphs of $\Phi_{\beta}$, $\Psi_{\beta}$ and $H_{\beta}$ for another recorded MCMC simulation. The results indicate that in the natural orientation (in the direction of time) an increase in $\Psi_{\beta}$ seems to be a trigger to an increase in the entropy and a decrease in the entropy seems to be a trigger to a decrease in $\Psi_{\beta}$. Roughly speaking, $\Psi_{\beta}$ ``pushes $H_{\beta}$ up'' and $H_{\beta}$ ``pushes $\Psi_{\beta}$ down''.

In summary, the central idea discussed here is that while entropy provides a measure of order/disorder of the system at a given configuration $\mathbf{X}^{(t)}$, Fisher information links these thermodynamical states through a path (\emph{Fisher curve}). Thus, Fisher information is a powerful mathematical tool in the study of complex and dynamical systems since it establishes how these different thermodynamical states are related along the evolution of the inverse temperature. Instead of knowing whether the entropy $H_{\beta}$ is increasing or decreasing, with Fisher information it is possible to know how and why this change is happening.

To test whether a system can recover part of its original coniguration after a perturbation is induced, we conducted another computational experiment. During a stable simulation, two kinds of perturbations were induced: 1) the value of the inverse temperature parameter was set to zero for the next consecutive 5 iterations; 2) the value of the inverse temperature parameter was set to the equilibrium value $\beta^{*}$ (solution of equation \ref{eq:Equilibrium}) for the next consecutive 5 iterations. 

\begin{figure}[!ht]
\begin{center}
\includegraphics[scale=0.6]{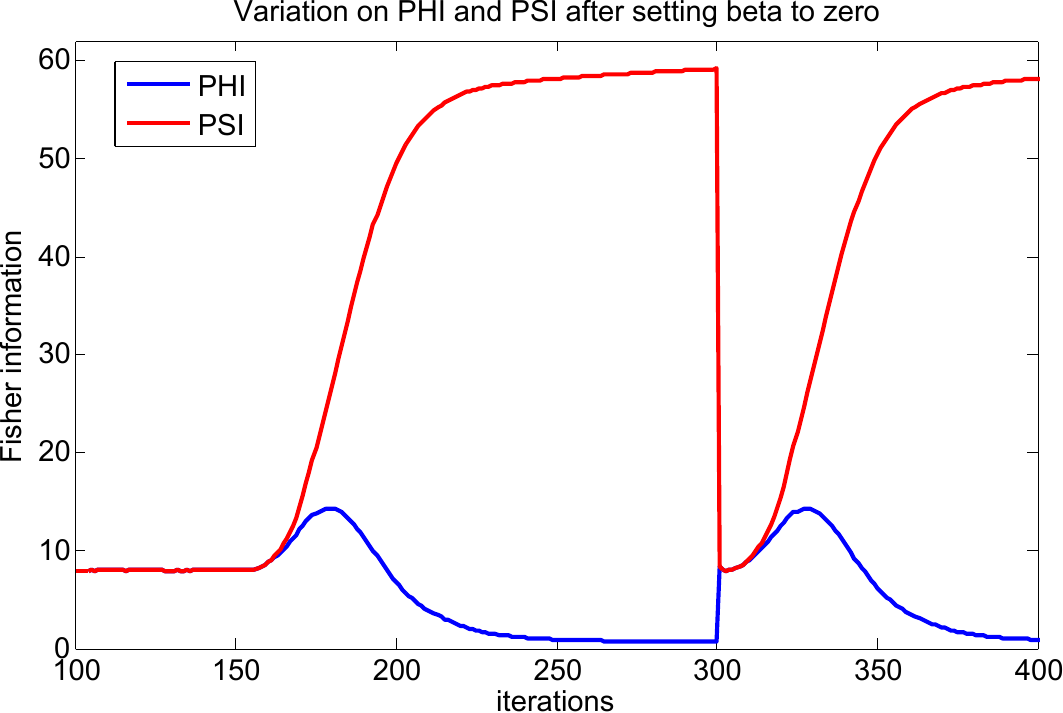}\vspace{1cm}
\includegraphics[scale=0.6]{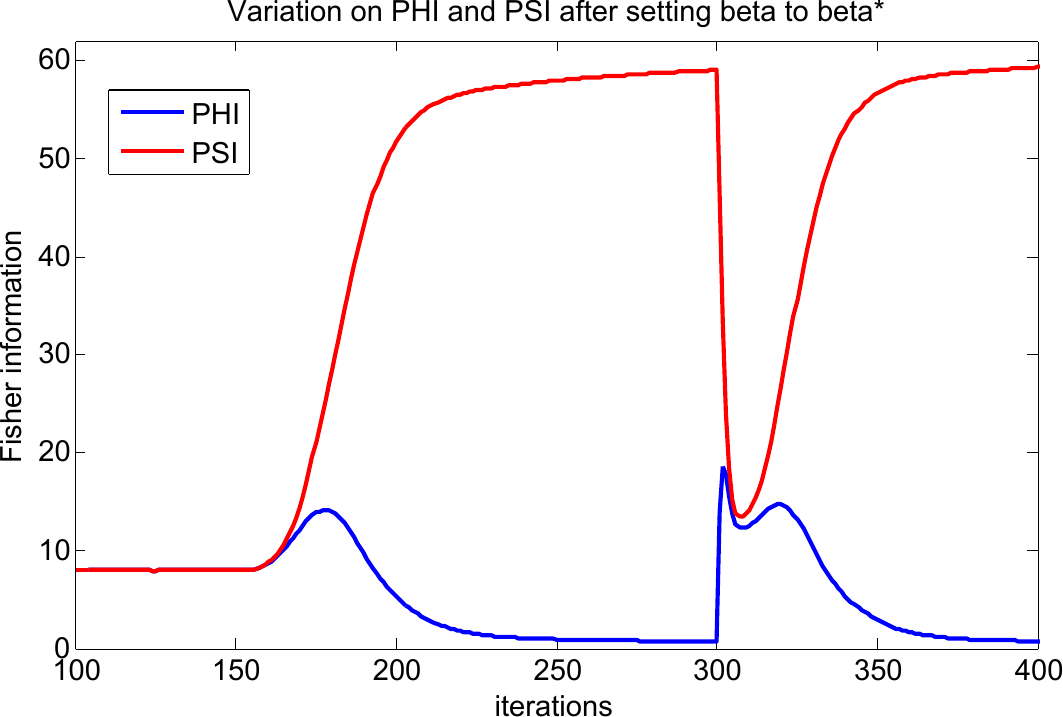}
\end{center}
\caption{
{\bf Disturbing the system to induce changes.} Variation on $\Phi_{\beta}$ and $\Psi_{\beta}$ after the system is disturbed by an abrupt change in the value of $\beta$. In the first image, the inverse temperature is set to zero. Note that $\Phi_{\beta}$ and $\Psi_{\beta}$ touch one another indicating that no residual information is kept, as if the simulation had been restarted from a random configuration. In the second image, the inverse temperature is set to the equilibrium value $\beta^{*}$. The results suggest that this kind of perturbation is not enough to remove all the information within the spatial dependence structure, allowing the system to recover a significant part of its original configuration after a short stabilization period.
}
\label{fig:res10}
\end{figure}

\begin{figure}[!ht]
\begin{center}
\includegraphics[scale=1]{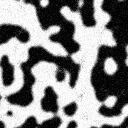}
\includegraphics[scale=1]{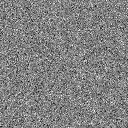}
\includegraphics[scale=1]{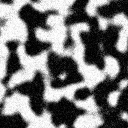}\\ \vspace{0.5cm}
\includegraphics[scale=1]{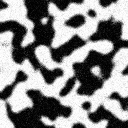}
\includegraphics[scale=1]{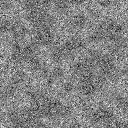}
\includegraphics[scale=1]{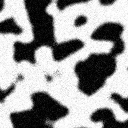}
\end{center}
\caption{
{\bf Sequence of outputs along the MCMC simulation before and after the system is disturbed.} The first row (when $\beta$ is set to zero) shows that the system evolved to a different stable configuration after the perturbation. The second row (when $\beta$ is set to $\beta^{*}$) indicates that the system was able to recover a significant part from its previous stable configuration.
}
\label{fig:res11}
\end{figure}

When the system is disturbed by seting $\beta$ to zero, the simulations indicate that the system is not successful in recovering components from its previous stable configuration (note that $\Phi_{\beta}$ and $\Psi_{\beta}$ clearly touch one another in the graph). When the same perturbation is induced but using the smallest of the two $\beta^{*}$ values (minimum solution of equation \ref{eq:Equilibrium}), after a short period of turbulence, the system can recover parts (components, clusters) of its previous stable state. This behavior suggests that this softer perturbation is not enough to remove all the information encoded within the spatial dependence struture of system, preserving some of the long-term correlations in data (stronger bonds), slightly remodeling the large clusters presented in the system. Figures \ref{fig:res10} and \ref{fig:res11} illustrate the results.

\section{Conclusions}

The definition of what is information in a complex system is a fundamental concept in the study of many problems. In this paper, we discussed the roles of two important statistical measures in isotropic pairwise Markov Random Fields composed of Gaussian variables: Shannon entropy and Fisher information. By using the pseudo-likelihood function of the GMRF model we derived analytical expressions for these measures. The definition of \emph{Fisher curve} as a geometric representation for the study and analysis of complex systems allowed us to reveal the intrinsic non-linear relation between these information-theoretic measures and gain insights about the behavior of such systems. Computational experiments demonstrates the effectiveness of the proposed tools in decoding information from the underlying spatial dependence structure of a Gaussian-Markov random field. Typical informative patterns in a complex systems are located in the boundaries of the clusters. One of the main conclusions of this scientific investigation concerns the notion of time in a complex system. The obtained results suggest that the relationship between Fisher information and entropy determines whether the system is moving forward or backward in time. Apparently, in the natural orientation (when the system is evolving forward in time), when $\beta$ is growing, that is, the temperature of the system is reducing, increase in Fisher information leads to an increase in the system's entropy and when $\beta$ is reducing, that is, the temperature of the system is growing, decrease in the system's entropy leads to decrease in Fisher information. Future investigations include the definition and analysis of the proposed tools in other Markov Random Field models, such as the Ising and Potts pairwise interaction models. Besides, a topic of interest concerns the investigation of minimum and maximum information paths in graphs to explore intrinsic similarity measures between objects belonging to a commom surface or manifold in $\Re^{n}$. We believe this study could bring benefits to some pattern recognition and data analysis computational applications.

\section{Acknowledgements}

The author would like to thank CNPQ (Brazilian Council for Research and Development) for the finantial support through the research grant number 475054/2011-3.

\bibliography{biblio}

\end{document}